\documentclass[journal]{new-aiaa}
\usepackage[utf8]{inputenc}
\usepackage{textcomp}

\usepackage{bm}
\usepackage{graphicx}
\usepackage{amsmath}
\usepackage[version=4]{mhchem}
\usepackage{siunitx}
\usepackage{longtable,tabularx}
\usepackage[table]{xcolor}
\definecolor{myblue}{rgb}{0, 0.23, 0.64}

\usepackage{hyperref}
\hypersetup{
    colorlinks=true,
    linkcolor=myblue,
    filecolor=magenta,
    citecolor=myblue,
    urlcolor=myblue,
}

\usepackage{footnpag}
\usepackage{mdframed}
\usepackage{mathtools}
\usepackage[ruled, vlined, linesnumbered]{algorithm2e}
\usepackage{subcaption}
\usepackage{multirow}
\usepackage{siunitx}
\usepackage{float}
\usepackage{alphalph}
\usepackage{etoolbox}
\usepackage{multicol}
\usepackage{booktabs}
\usepackage{threeparttable}
\usepackage{gensymb}
\usepackage{pdflscape}

\let\svthefootnote\thefootnote
\newcommand\freefootnote[1]{
  \let\thefootnote\relax
  \footnotetext{#1}
  \let\thefootnote\svthefootnote
}

\title{Concurrent Space-Based Observer Placement and Scheduling for Cislunar Space Domain Awareness}

\author{Dominic K. Amato\footnote{Master’s Student, Department of Mechanical, Materials and Aerospace Engineering. Student Member AIAA.} and Hang Woon Lee\footnote{Assistant Professor, Department of Mechanical, Materials and Aerospace Engineering; hangwoon.lee@mail.wvu.edu. Member AIAA (Corresponding Author).}}
\affil{West Virginia University, Morgantown, WV 26506}

\begin{document}

\freefootnote{This paper is a substantially revised and expanded version of paper AAS 25-691, presented at the 2025 AAS/AIAA Astrodynamics Specialist Conference, Boston, MA, August 10-14, 2025. It offers an updated formulation, an additional solution methodology, and new results.}

\maketitle

\begin{abstract}
As interest in cislunar space grows, so does the need for space domain awareness (SDA) capabilities to safely support missions. To enhance cislunar SDA capabilities, the use of space-based observers has been proposed. Prior research has explored optimizing observer placement but has not considered scheduling limitations. We propose a mixed-integer linear programming formulation that concurrently optimizes the placement and scheduling of space-based observers for cislunar SDA. Our work considers three aspects of observer scheduling: sensor tasking, communication, and charging. We utilize a scenario-based approach that generalizes observer placement for unknown targets while applying scheduling constraints to each scenario. However, this problem quickly becomes difficult for commercial solvers to handle as the number of scenarios grows. As an alternative, a column generation approach is employed to introduce well-performing schedules into the problem for each orbital slot. A comparative analysis is performed across multiple instances to evaluate the performance of column generation against that of a commercial solver. A case study is then conducted in which the performance of the observers' placement from the concurrent formulation is tested against a set of validation target sets and compared with the performance of two baselines in which placement and scheduling are performed sequentially.
\end{abstract}

\section*{Nomenclature}
{\renewcommand\arraystretch{1.0}
\noindent\begin{longtable*}{@{}l @{\quad=\quad} l@{}}
$B_\text{various}$ & observer battery related values\\
$b$ & auxiliary variable for observer battery capacity\\
$C$ & total power at the receiver\\
$D_\text{various}$ & observer data related values\\
$d$ & auxiliary variable for observer data capacity\\
$e$ & distance from transmitter to receiver\\
$\Lambda$ & Information matrix\\
$f$ & performance parameter\\
$H$ & measurement Jacobian\\
$\bm{h}$ & measurement vector\\
$\mathcal{I}$ & set of orbital slots, index $i$, cardinality $I$\\
$\mathcal{J}$ & set of targets, index $j$, cardinality $J$\\
$\mathcal{K}$ & set of schedules, index $k$, cardinality $K$\\
$\mathcal{K}'$ & set of schedules in restricted master problem, index $k$, cardinality $K'$\\
$N$ & number of observers\\
$n$ & observer placement decision variable\\
$Q$ & visibility condition of the relay satellite for observers\\
$q$ & communication decision variable\\
$R$ & measurement noise covariance matrix\\
$\bm{r}$ & position vector in synodic frame\\
$\mathcal{S}$ & set of scenarios, index $s$, cardinality $S$\\
$\mathcal{T}$ & set of time steps, index $t$, cardinality $T$\\
$u$ & observation decision variable\\
$V$ & visibility condition of target for observers\\
$\bm{v}$ & velocity in synodic frame\\
$W$ & visibility condition of the Sun from observers\\
$w$ & charging decision variable\\
$\bm{X}$ & state vector\\
$Z$ & objective value\\
$\alpha$ & azimuth bearing angle\\
$\bm{\gamma}_p$ & vector from observer to primary body\\
$\varepsilon$ & elevation bearing angle\\
$\lambda$ & schedule decision variable\\
$\bm{\rho}$ & vector from observer to target\\
$\sigma$ & standard deviation of measurement noise\\
$\Phi$ & state transition matrix\\
\end{longtable*}}

\section{Introduction}

With the current push from both governmental and commercial space entities to expand humanity's reach into space, cislunar space will begin to be populated with more objects in the coming years. These entities have already begun carrying out lunar missions \cite{Firefly, Odysseus, Athena, CAPSTONE} and have additional missions planned, including NASA's Artemis campaign, which plans to return humans to the Moon for the first time in decades \cite{Gateway1}. Missions to the cislunar domain are highly complex and require precise navigation, tracking, and coordination, given the chaotic nature of cislunar space, to ensure the safety of all involved. One aspect of cislunar operations that remains insufficiently developed is the available \textit{space domain awareness} (SDA) capability. There are currently limited capabilities to detect and track targets in cislunar space. The ability to do so depends on ground-based observers on Earth, but they come with many limitations. Due to the extreme distances involved with cislunar space, ground-based observers struggle to take accurate measurements of objects further in the domain. In addition, the region of space behind the Moon is constantly unable to be viewed by these observers. These limitations render ground-based observers ineffective for cislunar SDA.

To mitigate these issues, the use of space-based observers in cislunar orbits has been proposed. References~\cite{frueh2021cislunar, Frueh2021} provide early research into the potential of using space-based observers for tracking targets in cislunar space. Reference~\cite{Wilmer2022} explores using periodic cislunar orbits to view the Earth-Moon system's L1 and L2 Lagrange points for performing SDA on objects orbiting them. Reference~\cite{Fowler} evaluates the estimation performance of cislunar orbits using both a single observer and a pair of observers. Reference~\cite{Greaves2023} presents a robust tracking and navigation method for cislunar SDA operations and demonstrates optical-only SDA and accurate measurement classification. While these works demonstrate the potential of placing observers in cislunar orbits for improving our SDA capabilities, they do not leverage optimization techniques for the placement of the observers.

The placement of the observers in cislunar periodic orbits is one of the most impactful factors in optimizing their ability to track targets. Many prior works have investigated the optimization of the placement of space-based observers using various approaches and performance measures. Reference~\cite{Badura2023} explores using multiple different metrics for the objective of an observer placement optimization. Reference~\cite{Klonowski2024} designs cislunar constellation architectures that maximize target detectability while minimizing the fuel required for transfer trajectories. Reference~\cite{Visonneau2023} proposes an optimization formulation for a multi-satellite architecture that maximizes the observable area over the time horizon while minimizing both the number of satellites used and the total stability index of the satellites. Reference~\cite{Clareson2025} also optimizes the placement of a set of space-based observers through the use of an extended Kalman filter to determine the observers' ability to estimate the state of targets. Reference~\cite{Patel2024} formulates an observer placement optimization problem that minimizes the number of satellites required to meet spatio-temporally varying coverage demand. These works collectively demonstrate, in various ways, the benefits of optimizing the placement of space-based observers in cislunar space.

The tasking of which target each observer should observe at each time step is another important aspect of SDA that has been explored in prior literature. Reference~\cite{Eapen2024} optimizes a sensor tasking framework for a lunar-based observer and a space-based observer working together to observe the cislunar domain. References~\cite{Fahrner2022, Paul2025} perform the sensor tasking optimization for a set of observers, subsequently optimizing the architecture. Reference~\cite{Patel2025} approaches the problem of observer placement while concurrently optimizing the tasking of the observers for multi-target scenarios in order to provide maximal traffic coverage of cislunar orbits. Additionally, Reference~\cite{Shimane2025} proposes an approach for simultaneously designing a constellation of sensors and tasking the sensors using a time-expanded p-median problem.

While these works that focus on optimizing the placement and sensor tasking of observers in cislunar can be insightful into the types of orbits that are best at tracking objects in the domain, they neglect key aspects of real-world missions that affect the feasibility of their results. Previous research has been carried out under the assumption that observers can continuously track targets without regard to physical constraints on observers, such as their data and battery capacities over the mission lifespan. Considering these restrictions on the observers will improve the feasibility of the results obtained. This combination of considering the placement, sensor tasking, data, and battery of the observers transforms the problem into a \textit{concurrent placement and scheduling problem}. Scheduling problems that consider these aspects of satellites have been evaluated for Earth-observing satellites \cite{Pearl2025REOSSP, pearl_conf}, but have not been extended to cislunar observers. The aim of this paper is to incorporate these scheduling constraints into the placement optimization of space-based observers for cislunar SDA to improve the feasibility of the observers' placement.

The contributions of this paper are as follows. We present a mixed-integer linear programming (MILP) formulation for the concurrent placement and scheduling optimization of space-based observers in periodic cislunar orbits that considers the scheduling of the sensor tasking, data transmission, and charging of the observers. This proposed formulation is intended to better model realistic missions to cislunar space that must consider the limitations that data and battery restrictions have on the observers' capabilities. The placement is also generalized for unknown targets in our formulation by using a scenario-based optimization approach, in which the scheduling is optimized individually for different scenarios of target sets, but the placement is chosen by the overall performance across all scenarios. Using MILP allows for the guarantee that the solution obtained is globally optimal when the problem is solved to proven optimality. Due to the high complexity of the problem, as the solution space grows, the use of commercial MILP solvers to solve to optimality can become inefficient and even unusable at times. As this issue becomes more prevalent when considering the types of solution spaces that would be used in real mission design, an alternative way to solve the problem becomes necessary. Because of this, the use of column generation is explored to solve the concurrent placement and scheduling optimization of the observers. Column generation is a favorable method for our problem, in which the total number of schedules is far too large to consider them all. A comparative analysis is then performed between the performance of a commercial solver and that of the column generation approach designed in this work. The comparative analysis is done for a set of instances that varies the number of orbital slots, targets, and scenarios, to see where each solution method performs better. To demonstrate the value of our concurrent formulation, a case study is conducted to compare the performance of the concurrent optimization formulation with the performance of observers from two baselines in which observer placement and scheduling are performed sequentially. The results show that the concurrent formulation outperforms both baselines on average across a set of validation scenarios, justifying the need for concurrently optimizing the placement and scheduling of space-based observers. This paper builds on a preliminary version of this research \cite{Amato2025} by refining the formulation and offering new analyses of the solution method, the number of observers used, and the scheduling implementation.

This paper is organized as follows. First, the preliminaries for the models used in this research are presented in Sec.~\ref{sec:prelims}, including the dynamics, observation, and communication models. Next, Sec.~\ref{sec:formulation} outlines the MILP formulation designed for our observer placement and scheduling problem. Then, Sec.~\ref{sec:sol_method} discusses the column generation approach designed to solve our problem. Section~\ref{sec:comp_exp} then defines the parameters used for the computational experiments performed and presents the results of the solution method comparative analysis and the case study we perform. Lastly, Sec.~\ref{sec:conc} concludes the paper with a summary of the findings of our work and concepts for potential future expansions of this work.

\section{Preliminaries}
\label{sec:prelims}
This section first provides background on the circular restricted three-body problem (CR3BP) dynamics used to model the motion of objects in cislunar space in this work. Next, a description of how observability is addressed in this work is presented. Lastly, an outline of how communication is evaluated in this paper is provided.

\subsection{CR3BP Dynamics}
\begin{figure}[htbp]
    \centering
    \includegraphics[width=0.5\linewidth]{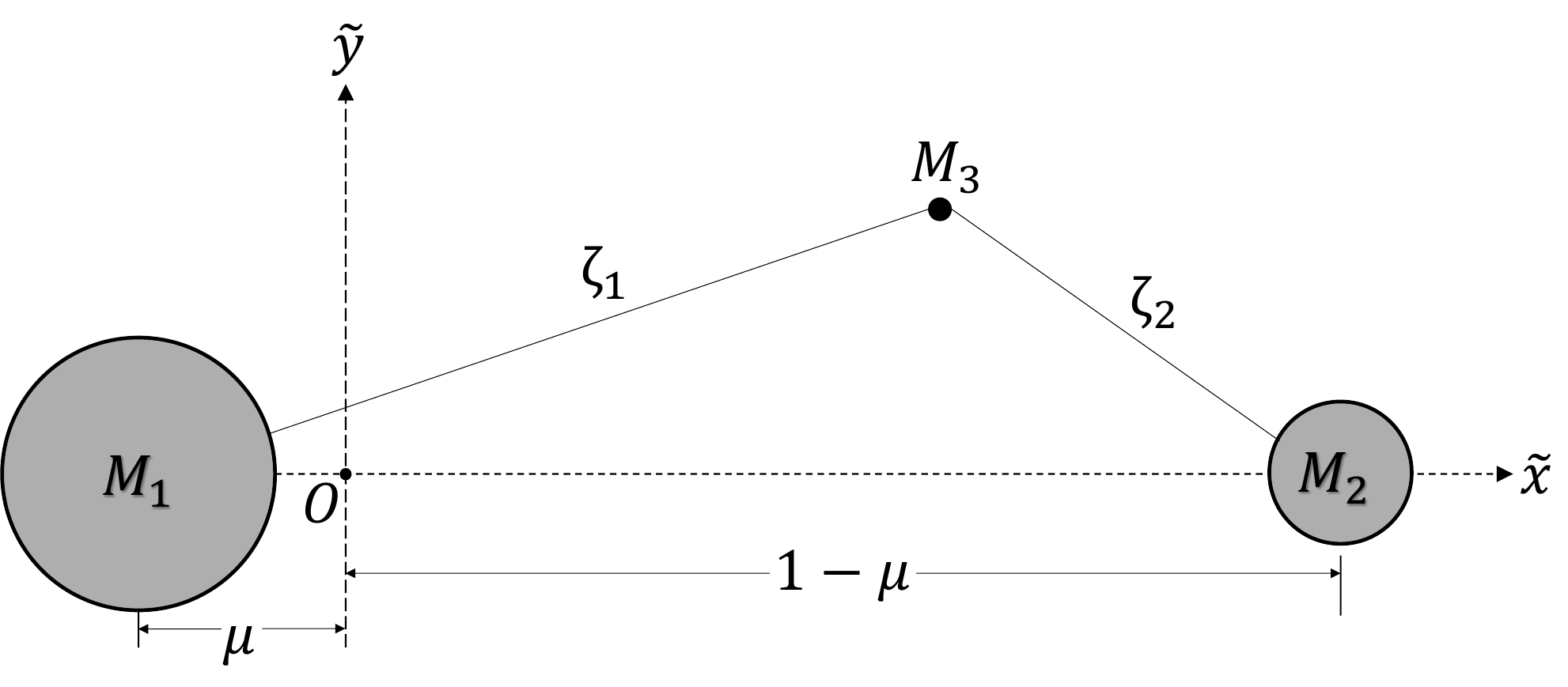}
    \caption{Synodic reference frame for the CR3BP.}
    \label{fig:synodic}
\end{figure}

For this paper, the CR3BP is used as the governing dynamics of all objects considered in the cislunar domain. For the CR3BP, a set of assumptions is made to simplify the dynamics of the system. The first assumption is that the Earth and Moon are considered the two primary bodies, and the mass of the third body, such as a satellite, is considered negligible compared to that of the other two. The other assumption made for the CR3BP is that the orbits of the two primary bodies are circular orbits about their common barycenter. A rotating reference frame $(\tilde{x},\tilde{y},\tilde{z})$, also referred to as a synodic frame, is used where the primary bodies lie on the $\tilde{x}$-axis, are rotating about the $\tilde{z}$-axis, and the $\tilde{y}$-axis completes the right-hand rule. The origin of the coordinate frame is at the barycenter of the primary bodies. This synodic frame is shown in Fig.~\ref{fig:synodic}.

In the synodic frame, the state of an object is represented using the components of its position $\bm{r} \in \mathbb{R}^3$ and velocity $\bm{v} \in \mathbb{R}^3$ in a vector $\bm{X}$
\begin{equation}
    \bm{X} = [\bm{r}^T,\bm{v}^T]^T
\end{equation}
where $\bm{r} = [x,y,z]^T$ and $\bm{v} = [\dot x, \dot y, \dot z]^T$. Unlike the two-body problem, the CR3BP has no closed-form analytical solution. Making these assumptions simplifies the dynamics of the system while still being able to capture the true governing dynamics using the following equations of motion:
\begin{subequations}
\label{EOM}
\begin{align}
    \Ddot{x} - 2\Dot{y} - x &= -\frac{(1-\mu)(x+\mu)}{\zeta_1^3} - \frac{\mu}{\zeta_2^3}(x-1+\mu) \label{EOM_x}\\
    \Ddot{y} + 2\Dot{x} - y &= -\frac{(1-\mu)}{\zeta_1^3}y - \frac{\mu}{\zeta_2^3}y \label{EOM_y}\\
    \Ddot{z} &= -\frac{(1-\mu)}{\zeta_1^3}z - \frac{\mu}{\zeta_2^3}z \label{EOM_z}
\end{align}
\end{subequations}
where $\mu$ is the nondimensionalized mass ratio, and $\zeta_1$ and $\zeta_2$ are the distances from the first and second primary bodies, respectively, to the third body, and are defined as:
\begin{subequations}
    \begin{align}
        \zeta_1 &= \sqrt{(x+\mu)^2+y^2+z^2} \\
        \zeta_2 &= \sqrt{(x-1+\mu)^2+y^2+z^2}
    \end{align}
\end{subequations}

Distance, mass, and time are all nondimensionalized for this work using the following equations, respectively.
\begin{subequations}
    \begin{align}
        l^* &= l_1 + l_2\\
        \mu &= \frac{M_2}{M_1+M_2}\\
        t^* &= \sqrt{\frac{(l^*)^3}{G(M_1+M_2)}}
    \end{align}
\end{subequations}
where $l_1$ and $l_2$ are the distances of each respective primary body from their barycenter, $M_1$ and $M_2$ are each respective primary body's mass, and $G$ is the gravitational constant. Table~\ref{tab:nondim} shows the resulting values used for the nondimensionalization of the Earth-Moon System \cite{jpl}. 

\begin{table}[htbp]
    \centering
    \caption{Nondimensionalized CR3BP units for Earth-Moon system.}
    \begin{tabular}{l c r}
    \hline
    \hline
     Nondimensionalized unit & Symbol  & Value  \\
     \hline
     Length unit (LU) & $l^*$   & \SI{389703}{km}\\
     Time unit (TU) & $t^*$   &   \SI{382981}{s}\\
     Mass ratio & $\mu$     & 0.0121506\\
     \hline
     \hline
    \end{tabular}
    \label{tab:nondim}
\end{table}

Furthermore, in the CR3BP for the Earth-Moon system, there are five equilibrium points, known as the Lagrange points and denoted $L_1-L_5$. In this work, we focus on orbits related to the $L_1$ and $L_2$ Lagrange points, as well as lunar-centered orbits.

\subsection{Observability}
\subsubsection{Measurement Model}
The measurement model used in this work is an angles-only model that considers the azimuth $\alpha$ and elevation $\varepsilon$ of the target with respect to the observer and the barycenter. An angles-only approach allows us to assume the use of low-cost optical sensors for this type of passive measurement. The relative position vector from the observer to the target $\bm{\rho}_\text{syn}$ is used to determine these angles and is calculated as follows.
\begin{equation}
        \bm{\rho}_\text{syn} = \left[
           x_{\text{tgt}},y_{\text{tgt}},z_{\text{tgt}} \right]^T - \left[ x_{\text{obs}},y_{\text{obs}},z_{\text{obs}}\right]^T
\end{equation}

For the calculation of the azimuth and elevation angles, the RSW body frame is used to align this work with satellite-based coordinate systems used in measurement models in the literature \cite{koblick, LVLH}. The R-axis is aligned with the radial position of the observer, the S-axis is aligned with the along-track displacement of the observer, and the W-axis is aligned with the vector normal to the observer's orbital plane. The rotation matrix to transform from the synodic to the RSW frame $R^{RSW}_\text{syn}$ is formed using the following equations:
\begin{subequations}
\label{eq:RSW}
\begin{align}
    \hat{\bm{R}} &= \frac{\bm{r}}{||\bm{r}||}\\
     \hat{\bm{W}} &= \frac{\bm{r} \times \bm{v}}{||\bm{r} \times \bm{v}||} \\
    \hat{\bm{S}} &= \hat{\bm{W}} \times \hat{\bm{R}}\\ 
    R^{RSW}_\text{syn} &= [\hat{\bm{R}},\hat{\bm{S}},\hat{\bm{W}}]^T
\end{align}
\end{subequations}
where $\hat{\bm{R}}$, $\hat{\bm{S}}$, and $\hat{\bm{W}}$ are unit column vectors. The relative position vector $\bm{\rho}$ is then transformed from the synodic frame to the RSW frame as shown in the following equation.
\begin{equation}
    \bm{\rho}_{RSW} = R^{RSW}_\text{syn} \bm{\rho}_\text{syn} \label{eq:rho_RSW}
\end{equation}

\begin{figure}[htbp]
    \centering
    \includegraphics[width=0.5\linewidth]{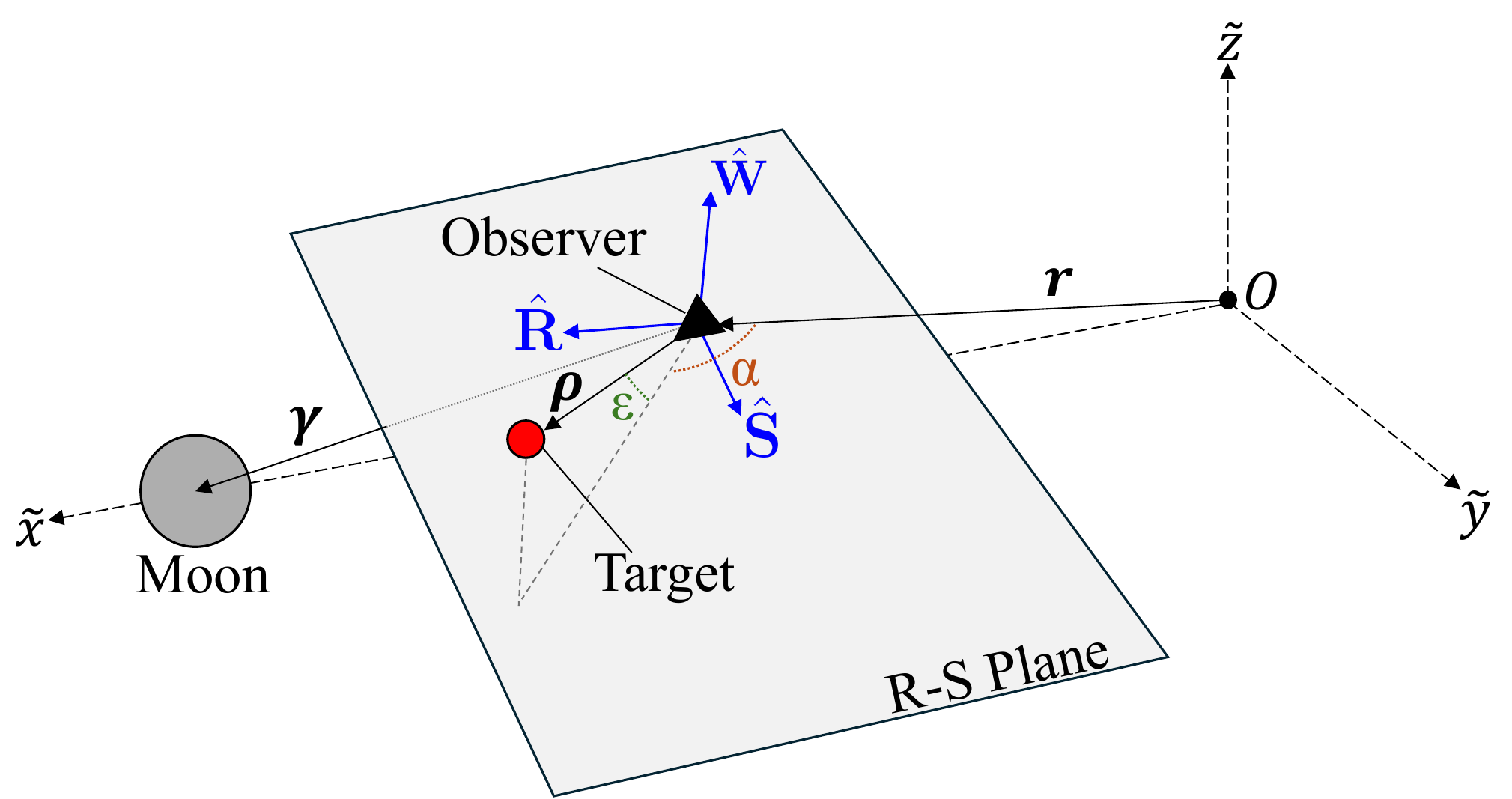}
    \caption{Depiction of the local body frame and angles-only measurement model.}
    \label{fig:meas_model}
\end{figure}

The angles for the measurement model are then calculated using the transformed $\bm{\rho}$ as shown in Eqs.~\eqref{eq:az} and \eqref{eq:el}, and form the measurement vector $\bm{h}$,
\begin{subequations}
    \begin{align}
    \alpha &= \operatorname{arctan2} \left(\rho_S,\, -\rho_R\right) \label{eq:az}\\
    \varepsilon &= \arcsin\left(\frac{\rho_W}{\rho_{RSW}}\right) \label{eq:el}\\
    \bm{h} &= [\alpha,\varepsilon]^T \label{eq:h}
    \end{align}
    \label{eq:angles}
\end{subequations}
where $\rho_{RSW}=\lVert\bm{\rho}_{RSW}\rVert$. The Jacobian of the measurement vector for the target $H$ is then calculated as follows:
\begin{equation}
    H=\frac{\partial{\bm{h}}}{\partial{\bm{X}_\text{tgt}}} \bigg|_{\bm{X}_{\text{tgt}},\bm{X}_{\text{obs}}} \label{eq:meas}
\end{equation}
where $\bm{X}_{\text{tgt}}$ and $\bm{X}_{\text{obs}}$ are the states of the target and observer, respectively. A depiction of the angles-only model with respect to the RSW coordinate frame is shown in Fig.~\ref{fig:meas_model}.

\begin{figure}[htbp]
    \centering
    \includegraphics[width=0.5\linewidth]{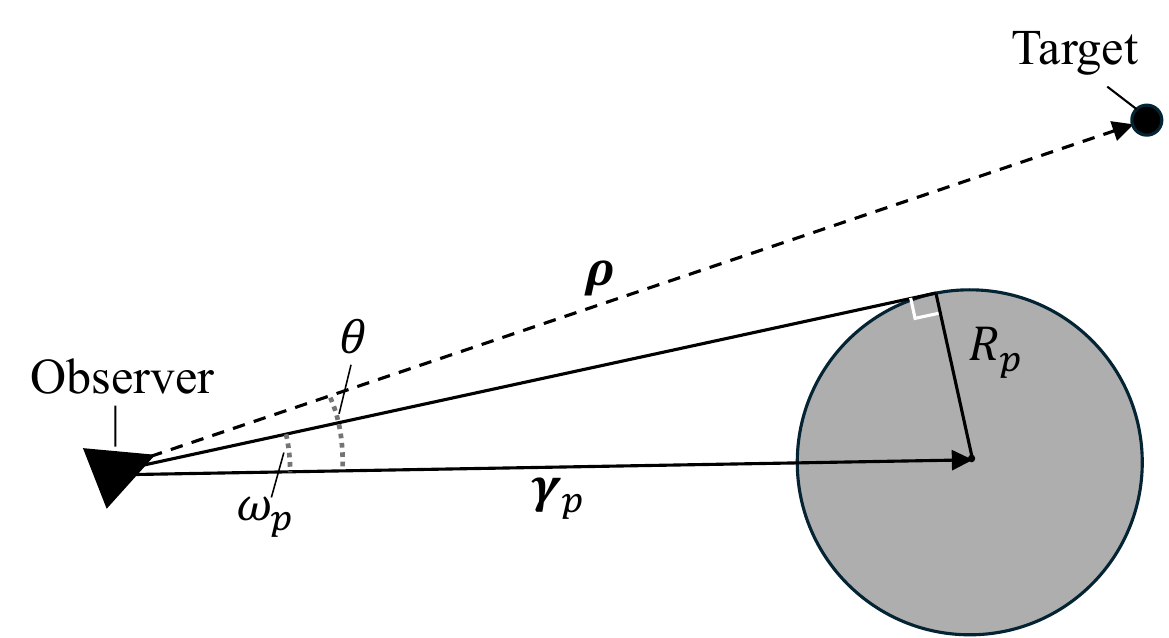}
    \caption{Angles used for occlusion constraint.}
    \label{fig:occlusion}
\end{figure}

Constraints are imposed to check for the geometric occlusion of the target with the Earth and the Moon. For each angular constraint, the angle $\theta$ between $\bm{\rho}_\text{syn}$ and $\bm{\gamma}_p$, the relative position vector from the observer to the primary body, must be greater than the occlusion angle $\omega_p$ for an observation to occur \cite{clareson2023opt, Clareson2025}. A target whose line of sight lies within the angular disk of a primary body is therefore not observable, whether it is behind or in front of that body. These angles are calculated as follows:
\begin{subequations}
    \begin{align}
        \theta &= \arccos \left( \frac{\bm{\rho_\text{syn}} \cdot \bm{\gamma}_p}{\lVert \bm{\rho}_\text{syn} \rVert  \lVert \bm{\gamma}_p \rVert} \right)\\
        \omega_p &= \arctan \left( \frac{R_p}{\sqrt{{\lVert \bm{\gamma}_p \rVert}^2 - R_p^2}} \right)
    \end{align}
\end{subequations}
where $R_p$ is the radius of the respective primary body and $\lVert \bm{\gamma}_p \rVert>R_p$. Figure~\ref{fig:occlusion} illustrates the angles used to check for occlusion with the primary bodies.

\subsubsection{Apparent Magnitude}
Apparent magnitude is also considered in this work as a determining factor for object visibility, representing the observed brightness of an object. The apparent magnitude considers a number of aspects pertaining to the observer, the target, and the environment. This work uses the model for apparent magnitude described by Vendl and Holzinger \cite{Vendl}. If the apparent magnitude of a target is less than the limiting magnitude of the observer $m_\text{lim}$, then the target is deemed visible to the observer. The apparent magnitude of the target is determined in comparison to the apparent magnitude of the Sun $m_\text{S} = -26.74$. The position of the Sun must be considered in the calculation of the apparent magnitude of the target. A simplified model of the position and motion of the Sun is considered for this work, in which the Sun is assumed to be in a coplanar clockwise circular orbit around the barycenter of the Earth-Moon system with an orbit radius of \SI{1}{AU} \cite{Patel2024}. The position of the Sun is initialized on the positive $\tilde{x}$-axis. The gravitational effects of the Sun are not considered in the dynamics of this work. The apparent magnitude of the target is a function of the solar phase angle $\Psi$, which is the angle between the relative position vector from the observer to the target $\bm{\rho}_\text{syn}$ and the relative position vector from the Sun to the target $\bm{\xi}$, and is calculated as follows.
\begin{equation}
    \Psi = \arccos \left( \frac{\bm{\rho}_\text{syn} \cdot \bm{\xi}}{\lVert \bm{\rho}_\text{syn} \rVert  \lVert \bm{\xi} \rVert} \right) \label{eq:solar_phase}
\end{equation}

The solar phase angle is then used to find the diffuse phase angle function $\phi$, calculated using Eq.~\eqref{eq:phi_diff}, which is in turn used in Eq.~\eqref{eq:app_mag} to determine the apparent magnitude of the target \cite{Vendl,Shell},
\begin{subequations}
    \begin{align}
        \phi &= \frac{2}{3\pi} [ \sin(\Psi) + (\pi - \Psi) \cos(\Psi) ]  \label{eq:phi_diff}\\
        m_\text{tgt} &= m_\text{S} - 2.5\log_{10}\left( \frac{\delta^2}{{\lVert \bm{\rho} \rVert}^2} \left[ \frac{a_\text{spec}}{4} + a_\text{diff} \phi(\Psi) \right] \right)  \label{eq:app_mag}
    \end{align}
\end{subequations}
where $\delta$ is the radius of the target, $a_\text{spec}$ is the specular reflectance, and $a_\text{diff}$ is the diffuse reflectance. The values we use for the constants in these equations and the parameters used for the motion of the Sun are shown in Table~\ref{tab:vis_mag}.
\begin{table}[htbp]
    \centering
    \caption{Apparent magnitude parameters \cite{Visonneau2023}.}
    \begin{tabular}{l r}
    \hline
    \hline
    Parameter & Value \\
    \hline
    Distance to sun, LU & 383.8821\\
    Angular velocity of Sun, \SI{}{rad/TU} & -0.9253\\
    Target radius, $m$ & 1\\
    Target specular reflectance & 0\\
    Target diffuse reflectance & 0.2\\
    Limiting magnitude & 15\\
    \hline
    \hline
    \end{tabular}
    \label{tab:vis_mag}
\end{table}

\subsubsection{Information Matrix}
The key focus of SDA is tracking noncooperative targets in space. This is conventionally done using a form of state estimation based on a set of measurements. Although common state estimation techniques such as Kalman filters are useful for determining an observer's ability to track a target \cite{Clareson2025,clareson2023opt}, they are incompatible with linear programming (LP) techniques. A popular substitute for these state estimation techniques is the use of the information matrix $\Lambda$, which quantifies the information provided by an observation and indicates how well the observation can constrain the state of an unknown target \cite{Stephens}.
 
A scalar function of the information matrix can serve as the metric for observability performance, as optimizing this scalar metric leads to improved observability at each time step, reducing the lower bound on local estimation uncertainty \cite{miga2022analytical}. This paper considers a discrete-time measurement system, in which the information matrix is calculated locally for each time step instead of in reference to a fixed time, and assumes identical, uncorrelated, zero-mean Gaussian noise on the azimuth and elevation measurements.
This form of the information matrix represents the local information contributed by an observation taken at time $t$ about the target state at the preceding time step $t-1$. Its trace serves as a heuristic measure of observation performance. The information matrix is calculated as follows:
\begin{equation}
    \Lambda(t,t-1) = \Phi(t,t-1)^T H_t^T R^{-1} H_t \Phi(t,t-1)
    \label{eq:FIM}
\end{equation}
where $\Phi(t,t-1)$ is the state transition matrix of the target from the preceding time step $t-1$ to the current time step $t$, $H_t$ is the measurement Jacobian mapping the target state to the observation space at time $t$, and $R$ is the measurement noise covariance matrix. The noise covariance matrix is a constant function and is defined in terms of the standard deviation of measurement noise $\sigma$ as follows \cite{miga2022analytical, Clareson2025}.
\begin{equation}
    R = \begin{bmatrix}
            \sigma^2 & 0\\ 0 & \sigma^2
        \end{bmatrix}
        \label{eq:R}
\end{equation}

\subsection{Communication}
Data transmission is another key aspect of this work, which incorporates operational constraints into the problem of cislunar space-based constellation design. Over the course of an observer's mission, it will gather a large amount of data on the targets it observes and tracks. The ability of observers to transmit the data they store is crucial to preventing them from running out of on-board storage, allowing them to continually observe targets over longer mission horizons. In addition, if smaller satellites such as CubeSats are used as observers, on-board processing might not be possible. As a result, it would require the observers to be able to communicate the data they gather back to a larger main satellite or ground stations on Earth in order to process the data on the observed targets. We assume the use of a relay satellite in a cislunar periodic orbit that the observers are able to transmit data to in order to alleviate their data storage.

Similar to checking whether an observer can view a target using parameters such as apparent magnitude, a link budget analysis is used to determine whether communication can occur between an observer and the relay satellite~\cite{SMAD}. The link budget analysis considers the gains and losses due to the hardware of both the transmitter and receiver, as well as the free space loss $L_\text{fs}$ due to the distance between the two, which is calculated using the following equation:
\begin{equation}
    L_\text{fs} = 10 \log_{10} \left[ {\left( \frac{4 \pi e \nu}{c} \right)}^2 \right] \label{eq:free_space}
\end{equation}
where $e$ is the distance between the transmitter and receiver, $\nu$ is the signal frequency, and $c$ is the speed of light. The parameters for the hardware of the observers and relay satellite, and their corresponding gains and losses, are consistent with the values used in similar literature on satellite communications in cislunar space \cite{Booth,Cetin2024,speretta2021,zhang}. The values used for the hardware parameters for the link budget analysis are provided in Table~\ref{tab:link_budget}.
\begin{table}[htbp]
    \centering
    \caption{Link budget parameters \cite{Booth,Cetin2024,speretta2021,zhang}.}
    \begin{tabular}{l r}
    \hline
    \hline
    Parameter & Value\\
    \hline
    Frequency & \SI{2.2}{GHz}\\
    Transmitter power & \SI{10}{W}\\
    Transmitter antenna diameter & \SI{0.25}{m}\\
    Transmitter antenna efficiency & \SI{55}{\%}\\
    Transmitter antenna gain & \SI{12.61}{dBi}\\
    Transmitter losses & \SI{1}{dB}\\
    Receiver antenna diameter & \SI{1.5}{m}\\
    Receiver antenna efficiency & \SI{55}{\%}\\
    Receiver antenna gain & \SI{28.17}{dBi}\\
    Receiver losses & \SI{3}{dB}\\
    System noise temperature & \SI{316.23}{K}\\
    Data transmission rate & \SI{160}{kbps}\\
    Required $E_b$/$N_o$ & \SI{2.5}{dB}\\
    Required link margin & \SI{3}{dB}\\
    \hline
    \hline
    \end{tabular}
    \label{tab:link_budget}
\end{table}

Using the parameters of the transmitter, the equivalent isotropic radiated power ($\mathrm{EIRP}$), which is a measure of the power radiated in the direction of the peak of the transmit antenna, is calculated using \cite{SMAD}:
\begin{equation}
    \mathrm{EIRP} = P_\text{T} + G_\text{T} - L_\text{T} \label{eq:EIRP}
\end{equation}
where $P_\text{T}$, $G_\text{T}$, and $L_\text{T}$ are the power, gains, and losses from the transmitter, respectively. The gains from the transmitter are dependent on its antenna's diameter $\Delta_T$ and signal frequency $\nu_T$, and are calculated as follows:
\begin{equation}
    G_\text{T} = 10 \log_{10} \left[ \eta\left(\frac{\Delta_\text{T} \pi \nu_\text{T}}{c}\right)^2 \right]
\end{equation}
where $\eta$ is the antenna efficiency. The $\mathrm{EIRP}$ can then be used alongside the parameters related to the receiver to determine the total power at the receiver $C$:
\begin{equation}
    C = \mathrm{EIRP} + G_\text{R} - L_\text{R} - L_\text{fs} \label{eq:tot_pow}
\end{equation}
where $G_\text{R}$ and $L_\text{R}$ are the gains and losses due to the receiver. The next step of the link budget analysis entails finding the ratio of carrier power to noise power spectral density $C/N_o$ using the following equation:
\begin{equation}
    C/N_o = C - \tau - k_b \label{eq:band_rat}
\end{equation}
where $\tau$ is the noise temperature of the receiver and $k_b$ is the Boltzmann constant. Using this ratio and the data transmission rate $R_b$, the signal-to-noise power ratio at the receiver $ E_b/N_0$ is determined as follows.
\begin{equation}
    E_b/N_o = C/N_o - R_b \label{eq:EbNo}\\
\end{equation}

Lastly, the link margin $\mathrm{LM}$ is found in Eq.~\eqref{eq:link_marg} by taking the difference between the calculated and required $E_b/N_o$.
\begin{equation}
    \mathrm{LM} = E_b/N_o -(E_b/N_o)_\text{req} \label{eq:link_marg}
\end{equation}

In order for the link to close and communication to occur, the calculated link margin must be greater than the receiver's required link margin. The line of sight from the observer to the relay satellite must also be clear of the Moon. All calculations in the link budget analyses are expressed in decibels.

\section{Optimization Formulation}
\label{sec:formulation}
This section presents the MILP formulation for the concurrent placement and scheduling optimization problem for a set of space-based observers. The section begins by defining the sets, parameters, and decision variables that compose the formulation. We then outline how the performance of the observers is maximized through the objective function and how the constraints are imposed on both the placement and actions of the observers. The set of scheduling, data, and battery constraints used for the observers in this work is inspired by the associated constraints from Refs.~\cite{pearl_conf, Pearl2025REOSSP}. The constraints are adapted for our cislunar SDA applications.  The full formulation for the concurrent observer placement and scheduling optimization is provided in Appendix~A.

\subsection{Sets, Parameters, and Decision Variables}
Table~\ref{tab:sets} provides a summary of the sets, parameters, decision variables, and auxiliary variables that apply to the observer placement and scheduling optimization problem discussed in this paper. The decision variables for the optimization include the placement and actions of the observers. All of the decision variables are defined as binary, such that for the placement decision variables, a value of one corresponds to an observer being placed in the associated orbital slot and a value of zero corresponds to no observer being placed in the associated orbital slot. These orbital slots represent both the orbit of the observers and the phase of the observer in the orbit at the epoch. Similarly, for the actions' decision variables, a value of one indicates the action is performed, while a value of zero indicates the action is not performed. In addition, auxiliary variables are incorporated into the formulation to track the data and battery capacities of the observers over the optimization period. 
\begin{table}[htbp]
\centering
    \caption{Definitions of sets, indices, and parameters.}
    \label{tab:sets}
    \centering
    \begin{tabular}{ll}
    \hline
    \hline
    Symbol & Description \\
    \hline 
    \multicolumn{2}{l}{Sets and Indices} \\
    $\mathcal{T}$   &Set of time steps (index $t$; cardinality $T$)\\
    $\mathcal{I}$ &Set of orbital slots considered for observers (index $i$; cardinality $I$)\\
    $\mathcal{J}$ &Set of targets in a scenario (index $j$; cardinality $J$)\\
    $\mathcal{S}$ &Set of scenarios considered (index $s$; cardinality $S$)\\
    \multicolumn{2}{l}{Constant Parameters} \\
    $D_\text{max} \ge 0$ & Maximum data capacity of observer\\
    $D_\text{obs} \ge 0$ & Data capacity used per observation\\
    $D_\text{comm} \ge 0$ & Data transmitted per communication\\
    $B_\text{max} \ge 0$ & Maximum battery capacity\\
    $B_\text{min} \ge 0$ & Minimum safe battery capacity\\
    $B_\text{obs} \ge 0$ & Battery used per observation\\
    $B_\text{comm} \ge 0$ & Battery used per communication\\
    $B_\text{charge} \ge 0$ & Battery charged per time step\\
    $B_\text{time} \ge 0$ & Battery used for standard operations per time step\\
    $f_{ijt}^s$ & Information gain from measurement made by observer $i$ observing target $j$\\ &in scenario $s$ at time step $t$\\
    $V_{ijt}^s$ & $\begin{cases}
       1, & \text{if target $j$ in scenario $s$ is visible to observer $i$ at time step $t$} \\
       0, & \text{otherwise}
   \end{cases}$\\
    $Q_{it}$ & $\begin{cases}
       1, & \text{if the relay satellite is visible to observer $i$ at time step $t$} \\
       0, & \text{otherwise}
   \end{cases}$\\
   $W_{it}$ & $\begin{cases}
       1, & \text{if the Sun is visible to observer $i$ at time step $t$} \\
       0, & \text{otherwise}
   \end{cases}$\\
   \multicolumn{2}{l}{Decision Variables} \\
    $n_{i}$ & $\begin{cases}
    1, & \text{if an observer is placed in orbital slot $i$} \\
    0, & \text{otherwise}
    \end{cases}$\\
    $u_{ijt}^s$ & $\begin{cases}
    1, & \text{if observer $i$ observes target $j$ in scenario $s$ at time step $t$} \\
    0, & \text{otherwise}
    \end{cases}$\\
    $q_{it}^s$ & $\begin{cases}
    1, & \text{if observer $i$ transmits to the relay satellite at time step $t$ in scenario $s$} \\
    0, & \text{otherwise}
    \end{cases}$\\
    $w_{it}^s$ & $\begin{cases}
    1, & \text{if observer $i$ charges at time step $t$ in scenario $s$} \\
    0, & \text{otherwise}
    \end{cases}$\\
     \multicolumn{2}{l}{Auxiliary Variables} \\
    $d_{it}^s \in [0,D_{\text{max}}]$ & Data capacity of observer $i$ at time step $t$ in scenario $s$\\
    $b_{it}^s \in [B_\text{min},B_{\text{max}}]$ & Battery capacity of observer $i$ at time step $t$ in scenario $s$\\
    \hline
    \hline
    \end{tabular}
\end{table}

\subsection{Objective Function}
The information matrix is used as a measure of the observability performance of the observers for tracking the targets. The natural logarithm of the trace of the information matrix is then used as the performance parameter for the objective function, represented as $f = \ln \operatorname{tr}(\Lambda)$ and referred to as the information gain of an observation. The trace grows with the inverse square of the observer-to-target range, so a few close approaches would otherwise dominate the sum; the logarithm limits this and keeps the objective coefficients of comparable magnitude. The trace exceeds one for every observation in our experiments, so $f$ is positive. Reference~\cite{Shen} outlines how the trace of the information matrix provides a measure of the information that a sensor can provide at a given time step, and utilizes summing the trace from multiple time steps. The objective function also considers the decision variable $u_{ijt}^s$, which determines if observer $i$ observes target $j$ in scenario $s$ at time step $t$. The performance is then summed over all time steps, targets, observers, and scenarios, and the optimization aims to maximize the following objective function.
\begin{equation}
    Z=\sum_{s \in \mathcal{S}} \sum_{i \in \mathcal{I}} \sum_{t \in \mathcal{T}} \sum_{j \in \mathcal{J}} f_{ijt}^s u_{ijt}^s  \label{opt:obj}
\end{equation}

\subsection{Observer Placement}
A set number of observers $N$ is placed into a set of cislunar orbital slots $\mathcal{I}$, which is determined by the decision variable $n_i$. These orbital slots represent a unique orbit as well as the position of the observer in the orbit at epoch. They are created using different initial conditions which are propagated for a given time horizon. The initial conditions we use for this work are taken from a JPL database \cite{jpl}. A phasing procedure is also performed on each orbital slot taken from the database to allow the observers to have different initial phases within each orbit and create additional orbital slots. Constraint~\eqref{opt:obs_place} limits the number of orbital slots filled to be equal to the number of observers \cite{Williams2026}.
\begin{equation}
    \sum_{i \in \mathcal{I}} n_i = N \label{opt:obs_place}\\
\end{equation}

The placement of the observers is only with respect to the set of orbital slots $\mathcal{I}$ and is not specified by the individual scenario $s$. The placement of the observers is generalized in this way to determine a single placement of each observer across all scenarios. All remaining constraints relate to the observers' schedules within each scenario. This type of formulation allows us to generalize the placement of the observers to attempt to simulate what would be unknown targets, while applying the deterministic constraints associated with scheduling to the individual scenarios.

\subsection{Visibility Time Windows}
The observers are able to perform three different actions: making observations of a target $u_{ijt}^s$, transmitting data to the relay satellite $q_{it}^s$, and charging their battery $w_{it}^s$. Each of the actions of the observers is constrained by their respective visibility time window (VTW). Constraints~\eqref{opt:obs_tar} restrict the ability for an observer to observe a target to only be possible when the target is visible to the observer, determined by $V_{ijt}^s$. Constraints~\eqref{opt:obs_comm} restrict the observer to only be able to transmit to the relay satellite when it is visible to the observer, determined by $Q_{it}$. Lastly, constraints~\eqref{opt:obs_sun} restrict the observer to only be able to charge when the Sun is visible to the observer, determined by $W_{it}$. The Sun is considered visible when the line of sight from the observer to the Sun is not blocked by the Earth or the Moon. All three actions are also only possible when an observer is placed into the respective orbital slot for the correlated action.
\begin{subequations}
\begin{align}
    & V_{ijt}^s n_i \ge u_{ijt}^s, &\quad \forall i \in \mathcal{I}, \forall j \in \mathcal{J}, \forall t \in \mathcal{T}, \forall s \in \mathcal{S} \label{opt:obs_tar}\\
    & Q_{it} n_i \ge q_{it}^s, &\quad \forall i \in \mathcal{I}, \forall t \in \mathcal{T}, \forall s \in \mathcal{S} \label{opt:obs_comm}\\
    & W_{it} n_i \ge w_{it}^s, &\quad \forall i \in \mathcal{I}, \forall t \in \mathcal{T}, \forall s \in \mathcal{S} \label{opt:obs_sun}
\end{align}
\end{subequations}

\subsection{Action Limits}
Constraints~\eqref{opt:act_obs} limit each observer to only be able to perform one action at a time. This restriction is considered to avoid potential issues with battery routing and overload and to incorporate sensor tasking by allowing each observer to observe only one target at a time. Constraints~\eqref{opt:comm_lim} also limit only one observer to transmit data to the relay satellite at a time. A link between an observer and the relay satellite requires the antenna of each to be pointed at the other, which would only be possible for one observer at a time.
\begin{subequations}
    \begin{align}
        & \sum_{j \in \mathcal{J}} u_{ijt}^s + q_{it}^s + w_{it}^s \leq 1, &\quad \forall i \in \mathcal{I}, \forall t \in \mathcal{T}, \forall s \in \mathcal{S} \label{opt:act_obs}\\
        & \sum_{i \in \mathcal{I}} q_{it}^s \leq 1, &\quad \forall t \in \mathcal{T}, \forall s     \in \mathcal{S} \label{opt:comm_lim}
    \end{align}
\end{subequations}

\subsection{Data and Battery Constraints}
Each observer has a set amount of data it can store on board, depending on the selected hardware. The data capacity of each observer is tracked throughout the optimization period using the auxiliary variable $d_{it}^s$, which represents the data capacity of an observer in orbital slot $i$ at time step $t$ in scenario $s$. The observers' data capacities are tracked to ensure that every decision made by the optimization is feasible by preventing an observer from attempting to store data when its capacity is full or from trying to transmit data when none is stored. Constraints~\eqref{opt:obs_data1} determine what the data capacity of an observer will be at the next time step based on the action performed at the current time step. Each action that affects an observer's data capacity has a correlated data value. $D_\text{obs}$ is the amount of data that an observer stores when it takes a measurement of a target, while $D_\text{comm}$ is the amount of data that an observer is able to transmit to the relay satellite. $D_\text{comm}$ is dependent on the data transmission rate of the observer as well as the time step size of the model. It is assumed that each observer starts with no data storage, shown in constraints~\eqref{opt:obs_data_init}.
\begin{subequations}
\begin{align}
   d_{i, t+1}^s &= d_{it}^s + D_\text{obs} \sum_{j \in \mathcal{J}}u_{ijt}^s - D_\text{comm} q_{it}^s, &\quad \forall i \in \mathcal{I}, \forall t \in \mathcal{T} \setminus\{T\}, \forall s \in \mathcal{S} \label{opt:obs_data1}\\
   d_{i, 1}^s &= 0, &\quad \forall i \in \mathcal{I}, \forall s \in \mathcal{S} \label{opt:obs_data_init}
\end{align}
\label{opt:obs_data}
\end{subequations}

The battery capacity of each observer is also tracked throughout the optimization period using the auxiliary variable $b_{it}^s$, which represents the battery capacity of an observer in slot $i$ at time step $t$ in scenario $s$. Similar to the data propagation constraint, constraints~\eqref{opt:obs_battery1} determine what the battery capacity of an observer will be at the next time step based on the action performed at the current time step. Unlike the observer's data capacity, all three possible actions affect the observer's battery capacity. $B_\text{charge}$ is the amount of battery that an observer is able to charge in a time step, $B_\text{obs}$ is the amount of battery that is used to observe a target, and $B_\text{comm}$ is the amount of battery used to transmit data to the relay satellite. Furthermore, an additional battery consumption parameter $B_\text{time}$ is incorporated to account for all other normal operational tasks of the observer and is also dependent on the time step duration. Lastly, in addition to a maximum battery capacity set by the assumed hardware, a nonzero minimum battery constraint is incorporated to provide a safety factor for the satellite's operation by preventing the observers from ever running out of power. It is also assumed that each observer starts with full battery capacity, shown in constraints~\eqref{opt:obs_battery_init}.
\begin{subequations}
\begin{align}
    b_{i, t+1}^s &= b_{it}^s + B_\text{charge} w_{it}^s - B_\text{obs} \sum_{j \in \mathcal{J}} u_{ijt}^s - B_\text{comm} q_{it}^s - B_\text{time}n_i, &\quad \forall i \in \mathcal{I}, \forall t \in \mathcal{T}\setminus\{T\}, \forall s \in \mathcal{S} \label{opt:obs_battery1}\\
    b_{i, 1}^s &= B_\text{max}, &\quad \forall i \in \mathcal{I}, \forall s \in \mathcal{S} \label{opt:obs_battery_init}
\end{align}
\label{opt:obs_battery}
\end{subequations}

\subsection{Full Formulation}
Given the objective function, constraints, decision variables, and auxiliary variables discussed previously, the full concurrent placement and scheduling optimization is formulated as follows:
\begin{subequations}
\begin{align}
    \max \quad & Z\nonumber\\
    \text{s.t.} \quad & \text{Constraints}~\eqref{opt:obs_place}-\eqref{opt:obs_battery}\nonumber\\
    & n_i \in \{0,1\}, &\quad \forall i \in \mathcal{I} \label{opt:bin_obs}\\
    & u_{ijt}^s \in \{0,1\}, &\quad \forall i \in \mathcal{I}, \forall j \in \mathcal{J}, \forall t \in \mathcal{T}, \forall s \in \mathcal{S} \label{opt:bin_observe}\\
    & q_{it}^s \in \{0,1\}, &\quad \forall i \in \mathcal{I}, \forall t \in \mathcal{T}, \forall s \in \mathcal{S} \label{opt:bin_obs_comm}\\
    & w_{it}^s \in \{0,1\}, &\quad \forall i \in \mathcal{I}, \forall t \in \mathcal{T}, \forall s \in \mathcal{S}\\
    &d_{it}^s \in [0,D_\text{max}], &\quad \forall i \in \mathcal{I}, \forall t \in \mathcal{T}, \forall s \in \mathcal{S} \\
    &b_{it}^s \in [B_\text{min},B_\text{max}], &\quad \forall i \in \mathcal{I}, \forall t \in \mathcal{T}, \forall s \in \mathcal{S}
    \label{opt:bin_obs_charge}
\end{align}
\label{opt:bins}
\end{subequations}

\section{Column Generation}
\label{sec:sol_method}
Due to the large number of variables and constraints, this problem quickly becomes too complex for even commercial solvers to solve effectively as the solution space grows. Several different alternative solution methods were explored for solving the problem. These include Benders decomposition, Lagrangian relaxation, and the method selected for this work, column generation. Column generation provides an efficient alternative that leverages the fact that it does not need to explicitly consider all possible variables during the solution process \cite{CG}. Instead, column generation starts with a small set of columns that provide feasible solutions, then iteratively adds new columns with the greatest potential to improve the solution. This iterative process is composed of two parts: a restricted master problem (RMP) and a set of pricing subproblems. This section outlines how the problem is reformulated for the column generation approach using Dantzig-Wolfe decomposition to form the RMP and pricing subproblems. We initialize a set of feasible columns to generate the restricted column set, which is used by the RMP. The LP-relaxed form of the RMP is then solved using a commercial solver. The dual variables of the RMP are then passed to the pricing subproblems, where new columns are generated based on their potential to improve the RMP's objective value. These new columns are then added to the restricted column set, and the process repeats until no improving column is found, the runtime limit is reached, or there is negligible improvement of the RMP objective value over multiple consecutive iterations. Once this process ends, the integrality constraints are re-imposed on the RMP, and the integral problem considering the final restricted column set is solved. The solution to this problem yields the optimum of the MILP problem restricted to the final column set, which serves as a heuristic solution for the original MILP problem.

\subsection{Dantzig-Wolfe Decomposition}
Using Dantzig-Wolfe decomposition, the original MILP formulation is reformulated into a master problem and a set of pricing subproblems \cite{Dantzig_Wolfe}. The columns for this problem represent schedules for individual orbital slots within a given scenario and are represented using $k \in \mathcal{K}_i^s$, where $ \mathcal{K}_i^s$ is the set of all feasible schedules for orbital slot $i$ in scenario $s$. We introduce the decision variable $\lambda_{ik}^s$ for which $\lambda_{ik}^s=1$ if schedule $k$  of orbital slot $i$ in scenario $s$ is chosen, and \num{0} otherwise. The scheduling variables are represented as $\tilde{\bm{u}}$, $\tilde{\bm{q}}$, and $\tilde{\bm{w}}$, each with a convex combination of extreme points $\tilde{\bm{u}}_k$, $\tilde{\bm{q}}_k$, and $\tilde{\bm{w}}_k$, as shown in Eqs.~\eqref{eq:conv1}--\eqref{eq:conv3}, which must be done for each orbital slot in each scenario. The addition of the tilde (\textasciitilde{}) indicates that these forms of the variables are fixed from the respective schedules, instead of being decision variables. For compactness, the time-indexed scheduling variables are represented in vector form over the set of time steps $\mathcal{T}$, which can be shown as $\tilde{\bm{u}}_{ijk}^s = [\tilde{u}_{ijkt}^s]_{t \in \mathcal{T}}$, $\tilde{\bm{q}}_{ik}^s = [\tilde{q}_{ikt}^s]_{t \in \mathcal{T}}$, and $\tilde{\bm{w}}_{ik}^s = [\tilde{w}_{ikt}^s]_{t \in \mathcal{T}}$.
\begin{subequations}
    \begin{align}
        &\tilde{\bm{u}}_{ij}^s = \sum_{k \in \mathcal{K}_i^s} \tilde{\bm{u}}_{ijk}^s \lambda_{ik}^s, &\forall i \in \mathcal{I}, \forall j \in \mathcal{J}, \forall s \in \mathcal{S} \label{eq:conv1}\\
        &\tilde{\bm{q}}_i^s = \sum_{k \in \mathcal{K}_i^s} \tilde{\bm{q}}_{ik}^s \lambda_{ik}^s, &\forall i \in \mathcal{I}, \forall s \in \mathcal{S} \label{eq:conv2}\\
        &\tilde{\bm{w}}_i^s = \sum_{k \in \mathcal{K}_i^s} \tilde{\bm{w}}_{ik}^s \lambda_{ik}^s, &\forall i \in \mathcal{I}, \forall s \in \mathcal{S} \label{eq:conv3}\\
        &\sum_{k \in \mathcal{K}_i^s}\lambda_{ik}^s = n_i, &\forall i \in \mathcal{I}, \forall s \in \mathcal{S} \label{eq:convex}
    \end{align}
\end{subequations}

Equation~\eqref{eq:convex} is the generalized convexity constraint for this problem. Alternatively to the conventional Dantzig-Wolfe decomposition, where the sum of columns for a single instance is equal to one, our convexity constraint has the sum of columns for each orbital slot within each scenario to be equal to the placement variable $n_i$. This constraint implies that columns should only be selected for orbital slots that have been filled, while the placement variable ensures that the same orbital slots are selected across all scenarios. Therefore, the schedules selected by the master problem correspond to the same orbital slots across all scenarios, guaranteeing the generalized placement of the observers as intended. Even though the LP-relaxation of the RMP allows $n_i$ to take fractional values, subsequently resulting in the sum of the column variables to be fractional as well, the fractional solutions remain consistent with respect to the observer placement due to the use of $n_i$ in the convexity constraint. 

The performance parameter $\kappa_{ik}^s$ for the RMP is represented as:
\begin{equation}
    \kappa_{ik}^s=\sum_{j \in \mathcal{J}} \sum_{t \in \mathcal{T}} f_{ijt}^s \tilde{u}_{ijkt}^s, \quad 
    \forall i \in \mathcal{I}, \forall k \in \mathcal{K'}_i^s, \forall s \in \mathcal{S}
\end{equation}
and the RMP is then formulated as follows:
\begin{subequations}
\begin{align}
    \max \quad & Z_\text{CG} = \sum_{s \in \mathcal{S}} \sum_{i \in \mathcal{I}} \sum_{k \in \mathcal{K'}_i^s}\kappa_{ik}^s \lambda_{ik}^s \label{cg:obj}\\
    \text{s.t.} \quad & \sum_{i \in \mathcal{I}} n_i = N\label{cg:place}\\
    & \sum_{i \in \mathcal{I}} \sum_{k \in \mathcal{K'}_i^s} \tilde{q}_{ikt}^s \lambda_{ik}^s \le 1, &\quad \forall t \in \mathcal{T}, \forall s \in \mathcal{S} \label{cg:comm}\\
    & \sum_{k \in \mathcal{K'}_i^s} \lambda_{ik}^s = n_i, &\quad \forall i \in \mathcal{I}, \forall s \in \mathcal{S}\label{cg:link}\\
    & \lambda_{ik}^s \in \{0,1\}, &\quad \forall k \in \mathcal{K'}_i^s, \forall i \in \mathcal{I}, \forall s \in \mathcal{S} \label{cg:var1}\\
    & n_i \in \{0,1\}, &\quad \forall i \in \mathcal{I} \label{cg:var2}
\end{align}
\end{subequations}
where $\mathcal{K'}_i^s$ represents the restricted column set for an orbital slot in a given scenario. Objective function~\eqref{cg:obj} aims to maximize the amount of total information gain from the set of schedules chosen. Constraints~\eqref{cg:place} and \eqref{cg:comm} are taken from Constraints~\eqref{opt:obs_place} and \eqref{opt:comm_lim} in the original formulation, respectively. The generalized convexity constraint for the restricted column set is enforced in constraints~\eqref{cg:link}, and constraints~\eqref{cg:var1} and \eqref{cg:var2} set the binary restrictions on the decision variables. The integrality constraints of the RMP are relaxed, and the problem is instead solved as an LP problem.

\subsection{Pricing Subproblems}
Let $\bm{\psi} \coloneqq \{\psi_{t}^s \colon t \in \mathcal{T}, s \in \mathcal{S}\}$ and $\bm{\beta} \coloneqq \{\beta_i^s \colon i \in \mathcal{I}, s \in \mathcal{S}\}$ denote the dual variables associated with constraints~\eqref{cg:comm} and \eqref{cg:link}, respectively. The dual variables are passed from the RMP to each of the pricing subproblems in order to determine the reduced cost. The pricing subproblems are solved separately for each possible orbital slot in each scenario, meaning $i$ and $s$ are parameterized in the subproblem formulation. The subproblems are formed using the remaining constraints from the original formulation that are not used in the RMP. Each of these constraints is correlated with the feasibility of the actions of a schedule for a given orbital slot in a specific scenario. Given an orbital slot in a given scenario, each pricing problem is formulated as follows:
\begin{subequations}
    \begin{align}
        \max \quad & \sum_{t \in \mathcal{T}}\sum_{j \in \mathcal{J}} f_{ijt}^su_{ijt}^s -\sum_{t \in \mathcal{T}} \psi_{t}^s q_{it}^s - \beta_i^s \label{pp:obj}\\
        \text{s.t.} \quad & V_{ijt}^s \ge u_{ijt}^s, &\quad \forall j \in \mathcal{J}, \forall t \in \mathcal{T}\\
        & W_{it} \ge w_{it}^s, &\quad \forall t \in \mathcal{T}\\
        & Q_{it} \ge q_{it}^s, &\quad \forall t \in \mathcal{T}\\
        & \sum_{j \in \mathcal{J}} u_{ijt}^s + q_{it}^s + w_{it}^s \leq 1, &\quad \forall t \in \mathcal{T}\\
        & d_{i, t+1}^s = d_{it}^s + D_\text{obs} \sum_{j \in \mathcal{J}}u_{ijt}^s - D_\text{comm}q_{it}^s, &\quad \forall t \in \mathcal{T}\setminus\{T\}\\
        & d_{i, 1}^s = 0\\
        & b_{i, t+1}^s = b_{it}^s + B_\text{charge} w_{it}^s - B_\text{obs} \sum_{j \in \mathcal{J}} u_{ijt}^s - B_\text{comm} q_{it}^s - B_\text{time}, & \quad \forall t \in \mathcal{T}\setminus\{T\}\\
        & b_{i, 1}^s = B_\text{max}\\
        & u_{ijt}^s \in \{0,1\}, &\quad \forall j \in \mathcal{J}, \forall t \in \mathcal{T}\\
        & q_{it}^s \in \{0,1\}, &\quad \forall t \in \mathcal{T}\\
        & w_{it}^s \in \{0,1\}, &\quad \forall t \in \mathcal{T}\\
        &d_{it}^s \in [0,D_\text{max}], &\quad \forall t \in \mathcal{T}\\
        &b_{it}^s \in [B_\text{min},B_\text{max}], &\quad \forall t \in \mathcal{T}
    \end{align}
\end{subequations}

Each pricing subproblem solves for the schedule with the greatest potential to improve the RMP. This is determined by the objective function~\eqref{pp:obj}, which considers both the performance in terms of information gain and the cost of the dual variables from the previous iteration of the RMP. The objective value obtained corresponds to the reduced cost. If the solution to a subproblem is found to have a positive reduced cost, then the column associated with this solution is added to the restricted column set for the next iteration of the RMP. If the solution to a subproblem has a nonpositive reduced cost, then no column is added to the RMP for that orbital slot in that scenario. If all of the subproblems for an iteration of the column generation algorithm produce solutions with a nonpositive reduced cost, no improving column has been found, and the iterative process is stopped. Each subproblem is solved with a time limit of \SI{5}{s} and the best solution found within that limit is used, so this stopping condition does not prove that no improving column exists.

\subsection{Column Initialization}
An initial set of feasible columns must first be produced in order to begin column generation. For our column initialization, a given number of feasible schedules is produced for each orbital slot in each scenario. These schedules are produced using the feasibility windows of each observer. At each time step, the possible actions are determined by the corresponding VTWs. From the set of possible actions at the time step, one action is chosen at random. A biased randomized approach is used, in which each possible observation action is given a higher probability of selection to increase the likelihood of improving the observers' performance. The new data and battery capacities are then checked using this chosen action. If the chosen action causes either the observer's data or battery to violate its capacity limits, that action is removed from the list of feasible actions for that time step, and a new action is chosen from the updated list. This is repeated at a given time step until a feasible action is found. This process is then repeated for all time steps, orbital slots, and scenarios until the desired number of schedules is produced for each orbital slot in each scenario pairing.

\section{Computational Experiments}
In this section, we present the computational experiments relating to the concurrent placement and scheduling optimization formulation created in this work. Section~\ref{sol_comp} compares the performance of the column generation approach defined in the previous section with the performance of a commercial MILP solver across multiple instances of solving the concurrent formulation. Section~\ref{case_study} presents the results of a case study showing the performance of the observers from the concurrent formulation against $20$ validation target sets. This performance is then compared with the performance of two alternative formulations that perform the placement and scheduling optimizations sequentially to demonstrate the need to consider these restrictions directly in the optimization. A second version of the scheduling comparative analysis is performed, in which the set of orbital slots considered for the observers' placement is first filtered based on the orbital slots' communication capabilities.
\label{sec:comp_exp}

\subsection{Experimental Setup}
The formulations presented in the previous sections are used for the computational experiments performed in the following sections. All experiments assume the placement and scheduling of $N=3$ observers. Table~\ref{tab:D_B_param} shows the data and battery parameters used for all experiments in this work. These parameters are based on values from other literature on small satellites and cislunar missions \cite{speretta2021,Cetin2024,Booth}. This hardware is chosen under the assumption that a constellation of smaller satellites would be used for SDA observations for both mission feasibility and financial reasons. Additionally, it is assumed that the maximum data and minimum battery limits are chosen with a small factor of safety in mind, as the actions chosen at the final time step do not consider these limits. We assume that these smaller satellites would be able to communicate their collected observation data to a larger relay satellite that could relay the data back to Earth. For all experiments, it is assumed that the relay satellite to which the observers transmit data is in an orbit similar to the near-rectilinear halo orbit selected for NASA's proposed Gateway space station. Although this project was canceled, the orbit is still chosen for this work as it is already a highly-studied orbit and was chosen by NASA due to the benefits of its reduced propulsion requirements and visibility of Earth allowing for communication with ground stations. 
\begin{table}[htbp]
\centering
\caption{Constant observer data and battery parameters \cite{speretta2021,Cetin2024,Booth}.}
\centering
\begin{tabular}{l r}
\hline
\hline
Parameter     & Value \\
\hline
$D_\text{max}$       & \SI{3}{GB} \\
$D_\text{obs}$       & \SI{19.15}{MB} \\
$D_\text{comm}$      & \SI{38.30}{MB} \\
$B_\text{max}$       & \SI{2000}{kJ} \\
$B_\text{min}$       & \SI{200}{kJ} \\
$B_\text{obs}$       & \SI{9.57}{kJ} \\
$B_\text{comm}$      & \SI{19.15}{kJ} \\
$B_\text{charge}$    & \SI{19.15}{kJ} \\
$B_\text{time}$      & \SI{0.96}{kJ} \\
\hline
\hline
\end{tabular}
\label{tab:D_B_param}
\end{table}

All experiments consider nine cislunar orbital families for both the orbits of the observers and the targets. They are the Northern Butterfly (NBF), Southern Butterfly (SBF), L1 Northern Halo (L1NH), L2 Northern Halo (L2NH), L1 Southern Halo (L1SH), L2 Southern Halo (L2SH), Distant Retrograde Orbit (DRO), Eastern Low Prograde (ELP), and Western Low Prograde (WLP). From the large set of initial conditions taken from the JPL database, certain constraints are applied to select a set of orbital slots suitable for our work. A maximum stability index of \num{1.2} is imposed for all orbital slots to ensure the observers can remain in the selected orbits for extended periods of time with a reasonable amount of station-keeping fuel. A maximum period of \SI{5}{TU} is also imposed to filter out extremely large orbits for computational efficiency. Additional orbital slots are then created by performing a phasing distribution on each current feasible orbital slot to form \num{10} orbital slots with initial conditions distributed at equal time intervals over one orbital period of the original orbital slot. A complete set of \num{19660} feasible orbital slots is produced using the process described above, with the number of orbital slots from each family shown in Table~\ref{tab:slots}. For all experiments, the targets considered are randomly sampled from a set of orbital slots formed by taking 150 orbital slots at evenly spaced indices from each of the cislunar families. This is done to give an equal chance of selection for the orbits of each family. A simulation horizon of \SI{5}{TU} is used for all experiments, with propagation time steps of \SI{0.001}{TU} and time steps of \SI{0.005}{TU} between observer actions. This simulation horizon is chosen to ensure that all orbital slots can complete at least one full orbit. A standard deviation of measurement noise of $3\degree$ is used for the calculation of the information matrix \cite{Patel2025}. The state transition matrix is computed as the product of the matrix exponentials of the Jacobian of the CR3BP equations of motion, each evaluated at the target's state and taken over one \SI{0.001}{TU} propagation step between the previous and current time step.
\begin{table}[htbp]
    \centering
    \caption{Number of feasible orbital slots from each cislunar family.}
    \begin{tabular}{l r}
    \hline
    \hline
    Cislunar family     & Number of orbital slots \\
    \hline
    Northern Butterfly     & 480\\
    Southern Butterfly     & 480\\
    L1 Northern Halo     & 190\\
    L2 Northern Halo    & 1460\\
    L1 Southern Halo     & 190\\
    L2 Southern Halo     & 1460\\
    Distant Retrograde Orbit     & 1900\\
    Eastern Low Prograde     & 4360\\
    Western Low Prograde     & 9140\\
    \hline
    \hline
    \end{tabular}
    \label{tab:slots}
\end{table}

All optimizations are computed on a platform equipped with an Intel Core Ultra 9 285 processor with a base frequency of 2.50 GHz (24 cores and 24 logical processors) and 128 GB of RAM. All optimizations are programmed in Python 3.14 with Gurobi 13.0.3, while all pre- and post-optimizations are programmed in MATLAB R2026a \cite{MATLAB}.

\subsection{Solution Method Comparative Analysis}
\label{sol_comp}
The performance of our column generation approach is compared with that of the commercial MILP solver, Gurobi Optimizer 13.0. Gurobi is a state-of-the-art solver for MILP problems and is widely used in similar optimization literature, which is why it is chosen as our baseline for comparison. The two solution methods are compared across a variety of instances, varying in both solution space and problem size, to explore where each method is best applied. Gurobi uses several techniques during its solution process, including presolving, heuristics, cutting planes, and branch-and-bound. The combination of these techniques is designed to improve the optimizer's performance by minimizing runtime and yielding better solutions.

The set of test instances is created by varying three of the cardinality sets: $\mathcal{I}$, $\mathcal{J}$, and $\mathcal{S}$. The set of instances is formed using all of the possible combinations created by selecting one value from each of the following sets: $\vert\mathcal{I}\vert \in \{135,180,225,315,450\}$, $\vert\mathcal{J}\vert \in \{3,5,10,20\}$, and $\vert\mathcal{S}\vert \in \{2,5\}$. The set of possible combinations creates \num{40} different instances, which are shown in Table~\ref{tab:instances}. The number of variables in these instances ranges from $1890135$ for the smallest instance to $54000450$ for the largest instance. In addition, the number of constraints for the instances ranges from $2162001$ for the smallest to $56255001$ for the largest.
\begin{table}[htb]
    \centering
    \caption{Test instances for solution method comparative analysis.}
    \begin{tabular}{c c c c @{\hspace{3em}} c c c c}
    \hline
    \hline
    Instance & $\vert\mathcal{I}\vert$ & $\vert\mathcal{J}\vert$ & $\vert\mathcal{S}\vert$ &
    Instance & $\vert\mathcal{I}\vert$ & $\vert\mathcal{J}\vert$ & $\vert\mathcal{S}\vert$ \\
    \hline
    1 & 135 & 3 & 2 & 21 & 225 & 3 & 5 \\
    2 & 135 & 5 & 2 & 22 & 225 & 5 & 5 \\
    3 & 135 & 10 & 2 & 23 & 225 & 10 & 5 \\
    4 & 135 & 20 & 2 & 24 & 225 & 20 & 5 \\
    5 & 135 & 3 & 5 & 25 & 315 & 3 & 2 \\
    6 & 135 & 5 & 5 & 26 & 315 & 5 & 2 \\
    7 & 135 & 10 & 5 & 27 & 315 & 10 & 2 \\
    8 & 135 & 20 & 5 & 28 & 315 & 20 & 2 \\
    9 & 180 & 3 & 2 & 29 & 315 & 3 & 5 \\
    10 & 180 & 5 & 2 & 30 & 315 & 5 & 5 \\
    11 & 180 & 10 & 2 & 31 & 315 & 10 & 5 \\
    12 & 180 & 20 & 2 & 32 & 315 & 20 & 5 \\
    13 & 180 & 3 & 5 & 33 & 450 & 3 & 2 \\
    14 & 180 & 5 & 5 & 34 & 450 & 5 & 2 \\
    15 & 180 & 10 & 5 & 35 & 450 & 10 & 2 \\
    16 & 180 & 20 & 5 & 36 & 450 & 20 & 2 \\
    17 & 225 & 3 & 2 & 37 & 450 & 3 & 5 \\
    18 & 225 & 5 & 2 & 38 & 450 & 5 & 5 \\
    19 & 225 & 10 & 2 & 39 & 450 & 10 & 5 \\
    20 & 225 & 20 & 2 & 40 & 450 & 20 & 5 \\
    \hline
    \hline
    \end{tabular}
    \label{tab:instances}
\end{table}

A different randomly sampled set of targets is used for each instance. The default parameters of Gurobi are used for all applications, except that an MIPGap tolerance of 0.5\% is used for both the strictly Gurobi solutions and the final integer RMP of the column generation approach. A runtime limit of \SI{3600}{s} is used for both solution methods. If Gurobi does not converge within the runtime limit, it returns the best incumbent solution it is able to obtain. The runtime limit is applied to the iterative process of column generation, which stops if the sum of the current runtime and the previous iteration's runtime exceeds the limit. This approximation attempts to leave time for the final integral solution to be produced after the iterative process is complete and accepts that there may be instances where the total runtime of the column generation instance is slightly under or over the runtime limit. To compare the performance of the two solution methods, a relative performance (RP) metric is defined as $RP = (Z_\text{CG}-Z_\text{G})/Z_\text{G}$, where $Z_\text{G}$ is the objective value of the solution obtained by Gurobi. If $RP>0$ for a given instance, the solution obtained by column generation outperforms the solution obtained by Gurobi for that instance, and if $RP<0$, the opposite is true.

\subsubsection{Solution Method Comparative Analysis Results}
The results for the instances with $\vert\mathcal{I}\vert \in \{135,180,225,315,450\}$ are shown in Table~\ref{tab:sol_comp_res}. The results shown for each instance chosen include the objective value for both solution methods; the MIPGap percentage, upper bound, and runtime for the solution found by Gurobi; the LP-relaxed RMP objective value; and the RP metric. The LP-relaxed RMP value is the optimum of the LP relaxation over the final restricted column set; it is an upper bound on $Z_\text{CG}$ but not on the optimum of the original MILP. The runtime is not presented for the column generation approach as every instance reaches the runtime limit. For each of these five values of $\vert\mathcal{I}\vert$, all eight pairings of $\vert\mathcal{J}\vert \in \{3,5,10,20\}$ and $\vert\mathcal{S}\vert \in \{2,5\}$ are shown, $40$ instances in total.

\begin{table}[htbp]
    \centering
    \small
    \caption{Solution method comparative analysis results.}
    \begin{threeparttable}
    \begin{tabular}{c c c r r r r r r r}
    \hline
    \hline
    &&& \multicolumn{4}{c}{\textbf{Gurobi}} & \multicolumn{2}{c}{\textbf{Column generation}} & \\
    \cmidrule(r){4-7} \cmidrule(l){8-9}
    $\vert\mathcal{I}\vert$ & $\vert\mathcal{J}\vert$ & $\vert\mathcal{S}\vert$ & \multicolumn{1}{c}{$Z_\text{G}$}\tnote{c} & MIPGap, \%\tnote{c} & UB\tnote{b,c} & Runtime,~\SI{}{min}\tnote{a,c} & \multicolumn{1}{c}{$Z_\text{CG}$} & LP-Relaxed RMP & RP, \% \\
    \midrule
     135 & 3 & 2 & 35418.43 & 0.47 & \textasciitilde & 42.98 & 26428.05 & 33831.60 & -25.38\\
      & 5 & 2 & 40257.97 & 0.50 & \textasciitilde & 41.13 & 28316.37 & 39116.40 & -29.66\\
      & 10 & 2 & 39537.90 & 3.20 & 40801.69 & - & 31268.59 & 39035.82 & -20.91\\
      & 20 & 2 & 46510.05 & 0.92 & 46936.73 & - & 35177.40 & 44836.48 & -24.37\\
      & 3 & 5 & 75057.50 & 39.28 & 104536.96 & - & 58507.11 & 78260.14 & -22.05\\
      & 5 & 5 & 102787.44 & 3.07 & 105941.23 & - & 70813.57 & 99418.53 & -31.11\\
      & 10 & 5 & 90297.08 & 56.13 & 140980.13 & - & 76918.94 & 96713.77 & -14.82\\
      & 20 & 5 & 64630.38 & 119.40 & 141802.20 & - & 88857.45 & 108840.90 & \textbf{37.49}\\
    \hline
     180 & 3 & 2 & 35932.87 & 1.28 & 36391.04 & - & 26001.88 & 35594.61 & -27.64\\
      & 5 & 2 & 42365.13 & 2.24 & 43312.56 & - & 31747.03 & 40054.97 & -25.06\\
      & 10 & 2 & 42860.42 & 2.69 & 44013.29 & - & 31995.99 & 42512.30 & -25.35\\
      & 20 & 2 & 53007.25 & 2.38 & 54266.34 & - & 39804.83 & 47039.11 & -24.91\\
      & 3 & 5 & 65127.05 & 66.55 & 108468.83 & - & 59546.03 & 80352.14 & -8.57\\
      & 5 & 5 & 43846.41 & 198.42 & 130845.63 & - & 71091.86 & 93875.90 & \textbf{62.14}\\
      & 10 & 5 & 54455.02 & 207.48 & 167437.30 & - & 75443.13 & 105608.77 & \textbf{38.54}\\
      & 20 & 5 & 40449.90 & 336.46 & 176547.50 & - & 83609.03 & 109310.56 & \textbf{106.70}\\
    \hline
     225 & 3 & 2 & 31078.17 & 16.35 & 36160.53 & - & 28080.70 & 33216.89 & -9.64\\
      & 5 & 2 & 40885.74 & 1.37 & 41444.14 & - & 27619.97 & 38908.18 & -32.45\\
      & 10 & 2 & 38955.47 & 5.12 & 40951.21 & - & 31064.26 & 38339.99 & -20.26\\
      & 20 & 2 & 47422.74 & 0.69 & 47750.81 & - & 36025.82 & 45829.90 & -24.03\\
      & 3 & 5 & 72749.46 & 46.29 & 106424.03 & - & 59237.24 & 79978.72 & -18.57\\
      & 5 & 5 & 95571.68 & 87.44 & 179136.35 & - & 71751.27 & 102589.50 & -24.92\\
      & 10 & 5 & 39773.02 & 315.65 & 165317.26 & - & 72011.71 & 95861.42 & \textbf{81.06}\\
      & 20 & 5 & 96674.67 & 121.64 & 214268.62 & - & 86905.09 & 110007.13 & -10.11\\
    \hline
     315 & 3 & 2 & 36811.74 & 0.47 & \textasciitilde & 49.03 & 26799.40 & 34969.43 & -27.20\\
      & 5 & 2 & 42948.75 & 19.98 & 51530.48 & - & 30797.69 & 40282.75 & -28.29\\
      & 10 & 2 & 41037.81 & 13.50 & 46576.84 & - & 33772.04 & 39202.94 & -17.71\\
      & 20 & 2 & 46837.25 & 11.08 & 52027.26 & - & 34767.68 & 46057.03 & -25.77\\
      & 3 & 5 & 71180.73 & 81.99 & 129544.18 & - & 60503.38 & 81986.26 & -15.00\\
      & 5 & 5 & 59422.34 & 205.36 & 181451.94 & - & 71743.55 & 91489.98 & \textbf{20.73}\\
      & 10 & 5 & 57850.96 & 251.31 & 203236.01 & - & 82907.16 & 98185.28 & \textbf{43.31}\\
      & 20 & 5 & 85542.75 & 179.67 & 239234.89 & - & 89879.26 & 110005.54 & \textbf{5.07}\\
    \hline
     450 & 3 & 2 & 42904.56 & 28.30 & 55048.24 & - & 31496.96 & 42537.93 & -26.59\\
      & 5 & 2 & 40424.60 & 31.39 & 53114.61 & - & 29536.05 & 40000.44 & -26.94\\
      & 10 & 2 & 41239.26 & 58.10 & 65198.19 & - & 36453.33 & 44374.71 & -11.61\\
      & 20 & 2 & 50540.22 & 16.19 & 58725.15 & - & 37342.98 & 46550.13 & -26.11\\
      & 3 & 5 & 73061.91 & 89.87 & 138723.79 & - & 63956.32 & 83802.05 & -12.46\\
      & 5 & 5 & 73033.05 & 151.42 & 183619.85 & - & 73763.31 & 91336.14 & \textbf{1.00}\\
      & 10 & 5 & 99841.94 & 103.83 & 203504.52 & - & 81329.19 & 99984.15 & -18.54\\
      & 20 & 5 & * & * & * & * & 74724.39 & 105688.14 & \textbf{N/A}\\
    \hline
    \hline
    \end{tabular}
    \begin{tablenotes}[flushleft]
    \footnotesize
     \item[a] A hyphen (-) indicates the trigger of the runtime limit of \SI{3600}{s} for Gurobi.
     \item[b] A tilde (\textasciitilde) indicates Gurobi was able to reach the optimality condition.
     \item[c] An asterisk (*) indicates that Gurobi ran out of memory while solving the problem, with a memory limit of \SI{100}{GB} per solve.
     \end{tablenotes}
     \end{threeparttable}
     \label{tab:sol_comp_res}
\end{table}

Gurobi reaches the MIPGap tolerance of 0.5\% within the runtime limit for three of the $40$ instances, and runs out of memory for one, the instance with $\vert\mathcal{I}\vert = 450$, $\vert\mathcal{J}\vert = 20$, and $\vert\mathcal{S}\vert = 5$. Of the remaining $39$ instances, Gurobi obtains the better solution for $30$ and the column generation approach for nine. With two scenarios, Gurobi obtains the better solution for all $20$ instances, by \num{10} to \num{32}\%, and its MIPGap after the runtime limit is at most \num{58}\%. With five scenarios, the outcome follows the MIPGap Gurobi reaches within the runtime limit: Gurobi obtains the better solution for every instance with a MIPGap below \num{119}\%, and the column generation approach for every instance with a MIPGap above \num{122}\%, by up to \num{107}\%. For the instances in which Gurobi converges, the column generation approach reaches \num{70} to \num{75}\% of the objective value obtained by Gurobi.

\subsection{Case Study}
\label{case_study}
A case study is performed on an instance from the prior comparative analysis. This case study is designed to test the performance of the optimized observer placement against new validation scenarios to assess whether the placement of the observers is generalized well for unknown targets as intended. We use instance~1 of Table~\ref{tab:instances} ($\vert\mathcal{I}\vert = 135$, $\vert\mathcal{J}\vert = 3$, $\vert\mathcal{S}\vert = 2$), for which Gurobi obtains an optimal solution within the MIPGap tolerance of 0.5\%. The placements are tested against $20$ validation sets, each consisting of a single scenario with three targets. The targets in each validation set are randomly selected from a pool of 100 orbital slots taken at evenly spaced indices from each of the nine orbital families. 

A comparative analysis is performed for this case study, comparing the performance against the validation sets of our concurrent formulation with two baselines in which the placement and scheduling are performed sequentially in a two-stage process. The first baseline is a completely unscheduled baseline that only considers the placement of the observers in the first stage of the optimization, without considering any sensor tasking or scheduling restrictions. The first stage of the unscheduled baseline is formulated as follows.
\begin{subequations}
    \begin{align*}
        \max\quad &Z\\
        \text{s.t.}\quad &\text{Constraints}~\eqref{opt:obs_place},\eqref{opt:obs_tar},\eqref{opt:bin_obs},\eqref{opt:bin_observe}
    \end{align*}
\end{subequations}

The second baseline incorporates sensor tasking into the placement optimization by allowing each observer to observe only one target at a time. This case is intended to represent the current state of the art in LP formulations for observer placement optimization for cislunar SDA. The first stage of the sensor tasking baseline is formulated as follows.
\begin{subequations}
    \begin{align*}
        \max\quad &Z\\
        \text{s.t.}\quad &\text{Constraints}~\eqref{opt:obs_place},\eqref{opt:obs_tar},\eqref{opt:bin_obs},\eqref{opt:bin_observe}\\
        &\sum_{j \in \mathcal{J}}u_{ijt}^s \le 1, \quad \forall i \in \mathcal{I}, \forall t \in \mathcal{T}, \forall s \in \mathcal{S}
    \end{align*}
\end{subequations}

In both cases, only the placement from the first stage is used; the schedules of the chosen observers are then determined in the validation sets as described below. This comparative analysis is designed to demonstrate the necessity of directly accounting for scheduling limitations in observer placement optimization. Both of the baselines are optimized using the same sets of orbital slots, scenarios, and targets as the concurrent formulation. The performance in each validation set is obtained in the same way for the concurrent formulation and both baselines: the placement is fixed, and the scheduling problem of our formulation, including the data and battery constraints and the limit of one observer communicating with the relay satellite per time step, is solved for the targets of that validation set. 

Table~\ref{tab:global_stats} shows the statistics comparing the performance of the concurrent formulation with the performances of the two baseline cases over the $20$ validation sets. The results show the concurrent formulation's mean improvement, median improvement, standard deviation of improvement, and mean improvement percentage over both baselines.
\begin{table}[htbp]
    \centering
    \caption{Scheduling comparative analysis statistics.}
    \begin{tabular}{l r r r r}
    \hline
    \hline
    Case     &  Mean improvement & Median improvement & Standard deviation & Mean improvement percent, \%\\
    \hline
    Unscheduled & 4522.71 & 4847.05 & 2668.42 & 38.91\\
    Sensor tasking & 8963.45 & 9714.02 & 3217.32 & 129.64\\
    \hline
    \hline
    \end{tabular}
    \label{tab:global_stats}
\end{table}

The concurrent formulation obtains the better result for $18$ of the $20$ validation sets against the unscheduled baseline and for all $20$ against the sensor tasking baseline. The mean improvements of $38.91\%$ and $129.64\%$ over the unscheduled and sensor tasking baselines, respectively, show that incorporating scheduling directly into the placement optimization is advantageous for producing high-quality feasible results.

\begin{figure}[htbp]
    \centering
    \includegraphics[width=0.7\linewidth]{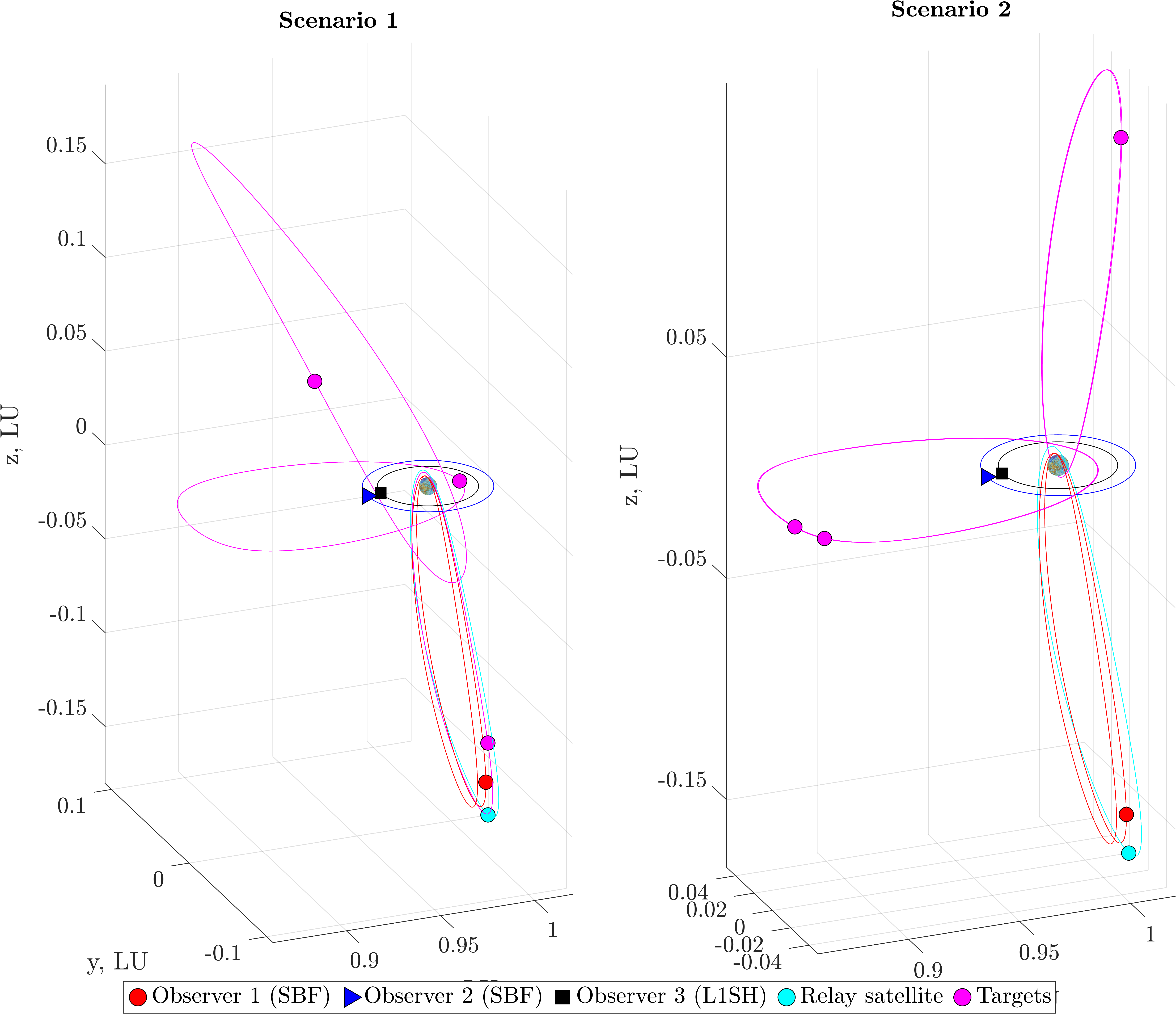}
    \caption{Observer placement results from concurrent optimization for both scenarios.}
    \label{fig:placement_w_targs}
\end{figure}

We choose to look in more detail at the results for one of the validation sets to clearly show the reason for the improvement of our concurrent formulation. Validation set~19 is chosen, as its improvement over the unscheduled baseline, \num{33.8}\%, is the closest to the mean improvement of \num{38.9}\%. The optimized placement of the observers for this instance is shown in Fig.~\ref{fig:placement_w_targs}, alongside the relay satellite and the targets from each of the two scenarios against which the observer placement is optimized. Observer~1 is placed into an SBF orbit, while observers~2 and~3 are placed into a DRO and a WLP orbit, respectively.

\begin{figure}[htbp]
    \centering
    \includegraphics[width=0.7\linewidth]{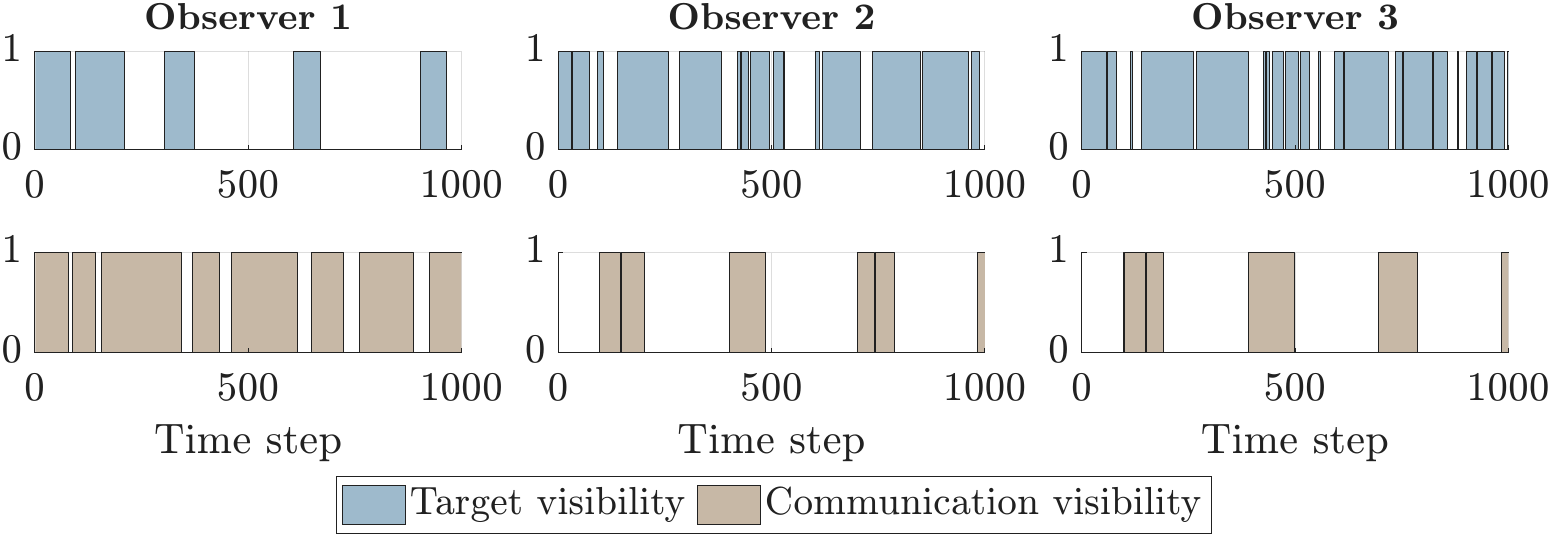}
    \caption{Observation and communication visibility of observers for the chosen validation set.}
    \label{fig:con_vis}
\end{figure}

\begin{figure}[htbp]
    \centering
    \includegraphics[width=0.8\linewidth]{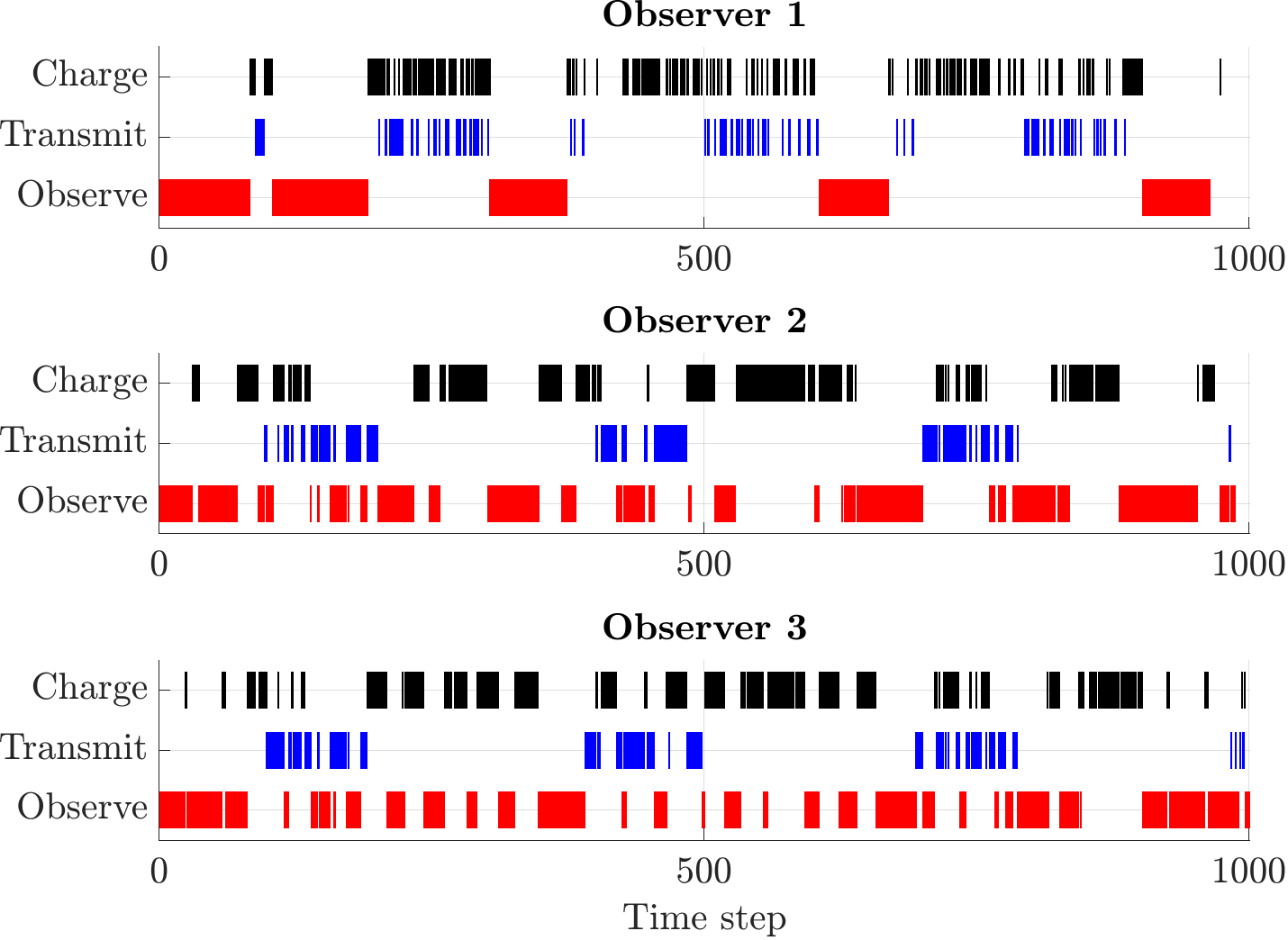}
    \caption{Optimal schedules of observers for the chosen validation set.}
    \label{fig:con_sched}
\end{figure}

Figure~\ref{fig:con_vis} shows which observers are able to make observations and communicate at each time step during the chosen validation set. Observer~1 can communicate with the relay satellite for most of the mission horizon but sees the targets only in five intervals, while observers~2 and~3 see the targets for most of the horizon and can communicate with the relay satellite in three intervals each. The optimal schedules in Fig.~\ref{fig:con_sched} reflect this division: observers~2 and~3 observe for most of the horizon and transmit their stored data during their communication windows, while observer~1 observes whenever the targets are visible and transmits in between. The placement thus combines one observer with relay access for most of the mission horizon and two observers with long target visibility.

\begin{figure}[htbp]
    \centering
    \includegraphics[width=0.8\linewidth]{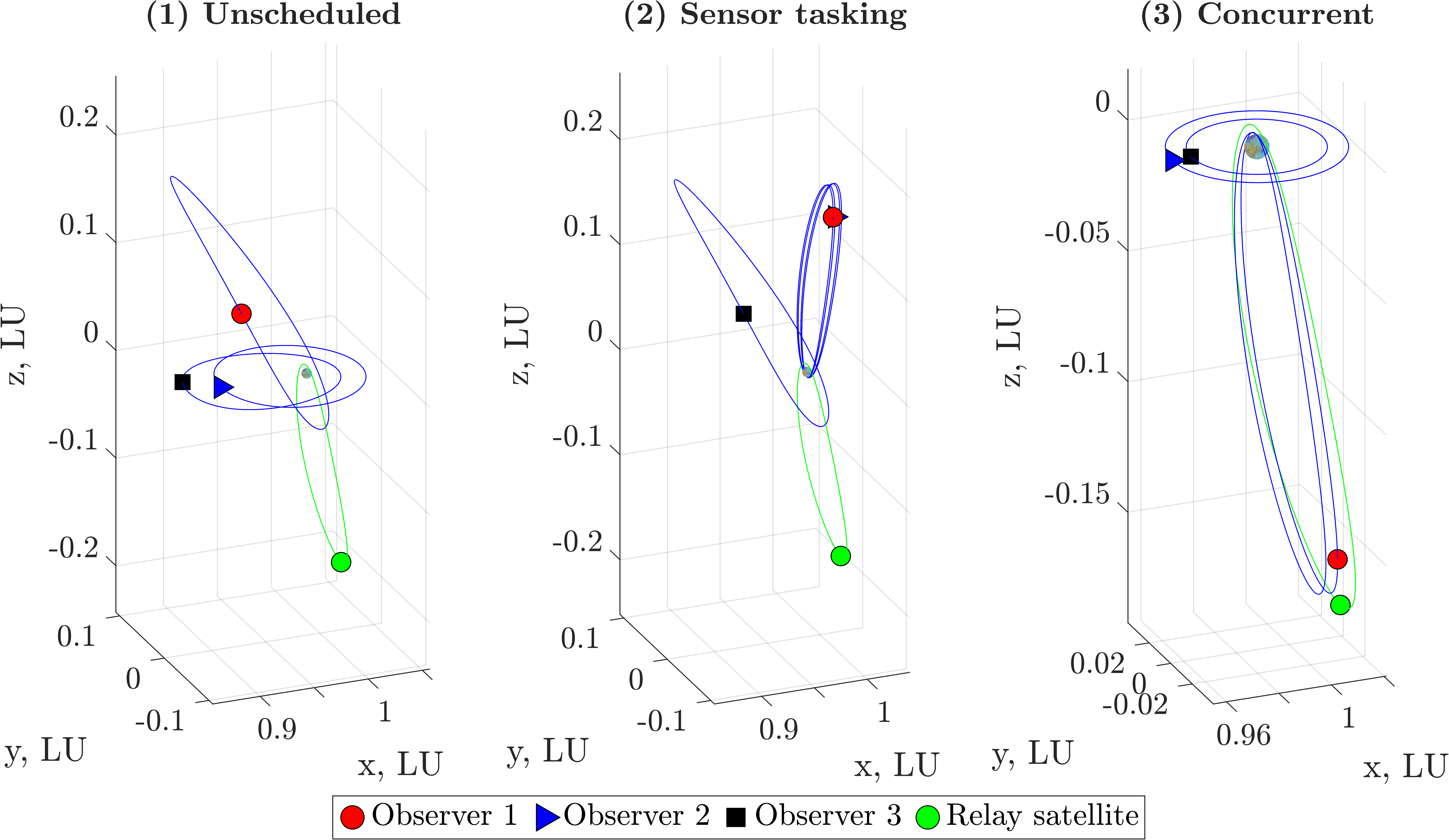}
    \caption{Plots of observer placements for the three cases.}
    \label{fig:placement}
\end{figure}

Figure~\ref{fig:placement} shows the optimized placement of the observers from the two baselines alongside the observer placement from the concurrent formulation. The unscheduled baseline places one observer into an L1NH orbit and two into WLP orbits, while the sensor tasking baseline places two observers into NBF orbits and its third into the same L1NH orbital slot as the unscheduled baseline. The initial conditions for the orbital slots chosen for all three cases of this comparative analysis are provided in Table~\ref{tab:init_cond} in Appendix~B.

The performance of the baselines against the chosen validation set is then evaluated to compare with that of the concurrent formulation. Table~\ref{tab:sched_comp_results} contains the results of the comparative analysis between the different observer placement formulations. The results include the orbit family to which each observer is assigned, the performance of each case in the chosen validation set, and the percentage improvement in performance over the baseline case. This percentage improvement is found using $(Z-Z_{B1})/Z_{B1}$, where $Z$ and $Z_{B1}$ are the objective value in the chosen validation set of the case being compared and the unscheduled baseline, respectively.
\begin{table}[htbp]
    \centering
    \caption{Scheduling comparative analysis results.}
    \begin{tabular}{l c c c r r}
    \hline
    \hline
    & \multicolumn{3}{c}{Orbit family}\\
    \cmidrule(r){2-4}
    Case     &  Observer 1 & Observer 2 & Observer 3 & Performance & Improvement over unscheduled, \%\\
    \hline
    Unscheduled & L1NH & WLP & WLP & 12109.04 & -\\
    Sensor tasking & NBF & NBF & L1NH & 6965.76 & -42.47\\
    Concurrent & SBF & DRO & WLP & 16206.13 & 33.84\\
    \hline
    \hline
    \end{tabular}
    \label{tab:sched_comp_results}
\end{table}

\begin{figure}[htbp]
    \centering
    \begin{subfigure}[b]{0.7\textwidth}
        \centering
        \includegraphics[width=\linewidth]{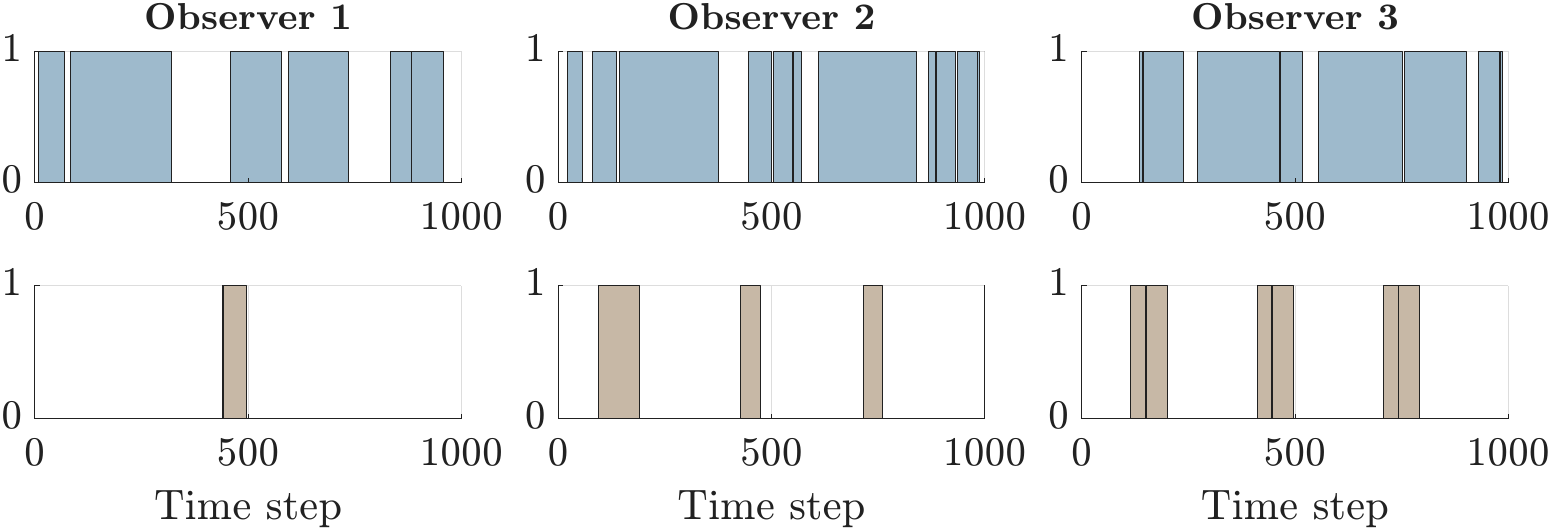}
        \caption{Unscheduled}
	    \label{fig:non_sched_vis}
    \end{subfigure}
    \begin{subfigure}[b]{0.7\textwidth}
        \centering
        \includegraphics[width=\linewidth]{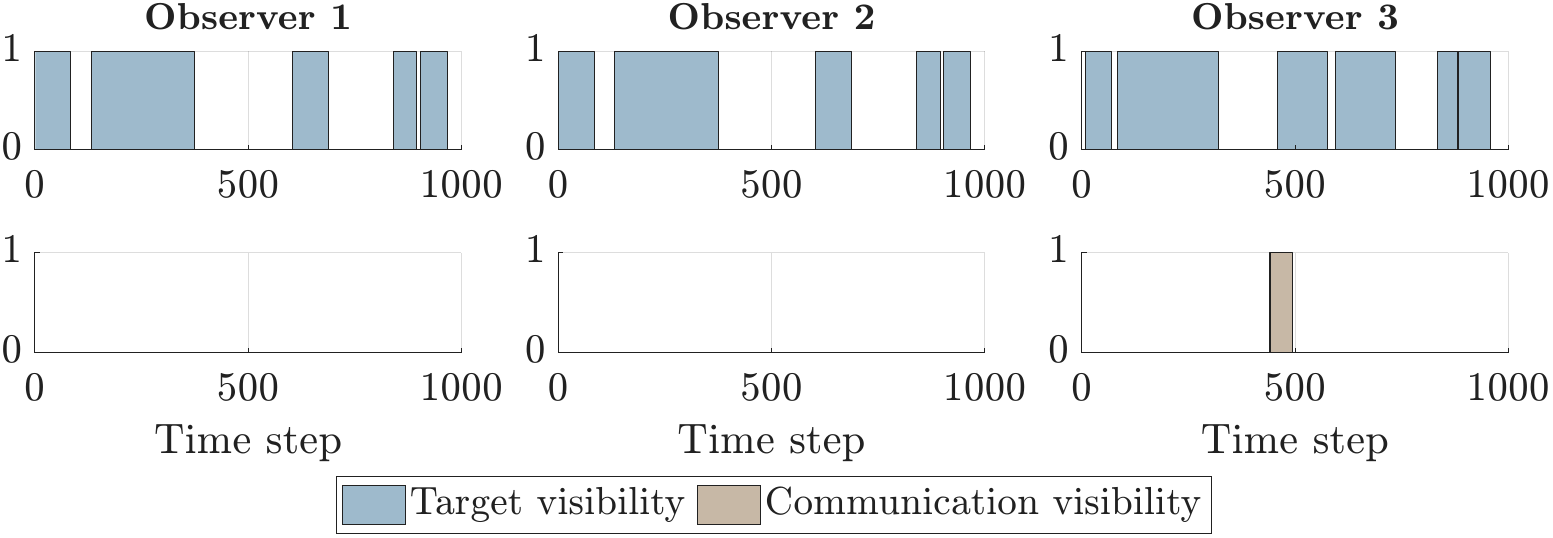}
        \caption{Sensor tasking}
	  \label{fig:sens_task_vis}
    \end{subfigure}
    \caption{Observation and communication visibility of the observers for the chosen validation set.}
    \label{fig:comp_vis}
\end{figure}

The sensor tasking baseline performs \num{42.47}\% worse than the unscheduled baseline. Two of its observers are placed into NBF orbital slots that have no communication visibility to the relay satellite at any time step, and its third observer, in the L1NH orbital slot, can communicate only in one short interval, as shown in Figs.~\ref{fig:non_sched_vis} and \ref{fig:sens_task_vis}; $20$ of the $135$ orbital slots of this instance have no communication visibility. The unscheduled baseline shares this L1NH observer, but its two WLP observers can communicate with the relay satellite in three short intervals each. Because the placement stage of both baselines does not consider data transmission, it selects orbital slots for their visibility of the targets regardless of their ability to transmit data. Our formulation outperforms the unscheduled and sensor tasking baselines by \num{33.84}\% and \num{132.65}\%, respectively, in this validation set, justifying the need to consider scheduling constraints concurrently in placement optimization rather than sequentially.

\begin{figure}[htbp]
    \centering
    \includegraphics[width=0.7\linewidth]{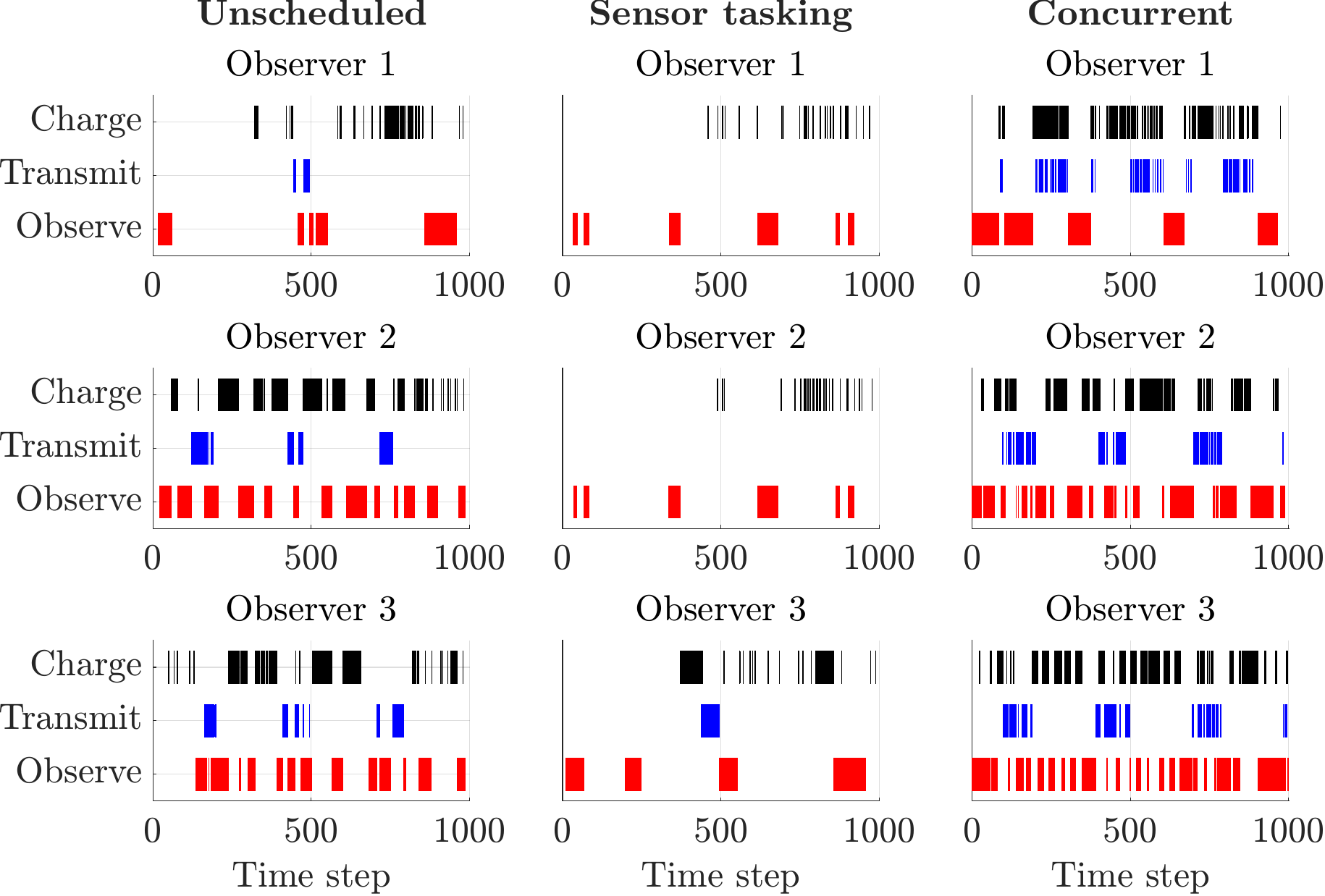}
    \caption{Optimal schedules of the three cases for the chosen validation set.}
    \label{fig:schedules}
\end{figure}

\begin{figure}[htbp]
    \centering
    \includegraphics[width=0.7\linewidth]{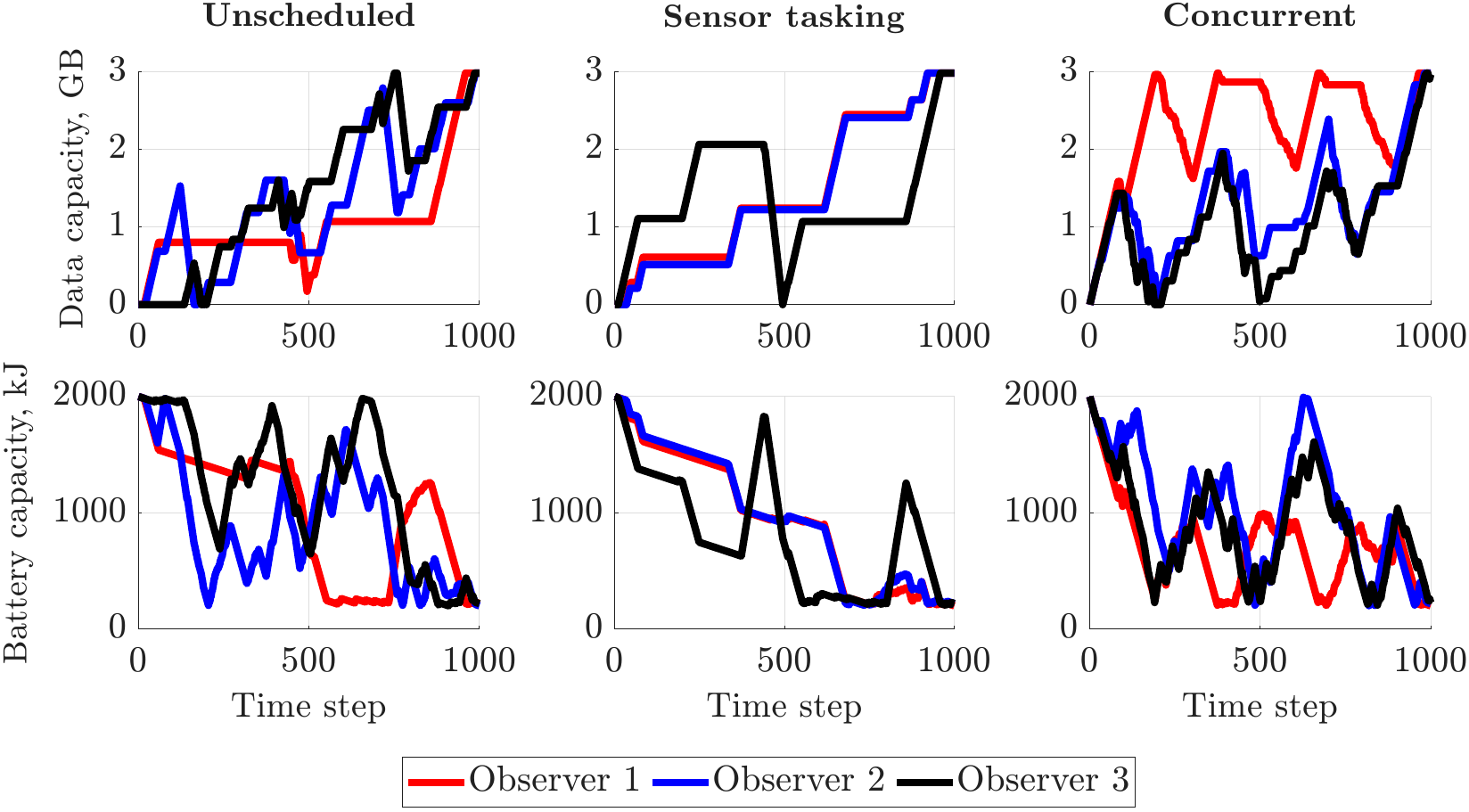}
    \caption{Data and battery capacities of observers during optimal schedules for the three cases.}
    \label{fig:capacities}
\end{figure}

The optimal schedules for each of the three cases in the chosen validation set are shown in Fig.~\ref{fig:schedules}. The observers of the sensor tasking baseline transmit in only one interval, and those of the unscheduled baseline only during their few short communication windows. Their data capacities in Fig.~\ref{fig:capacities} therefore mostly increase, in steps whenever the targets are visible, until they reach the data capacity limit, after which no further observations are possible. The observers of the concurrent formulation transmit their data repeatedly, so that their data capacities rise and fall throughout the mission horizon. They also charge more frequently than the observers of the sensor tasking baseline, providing the battery energy needed for both observations and transmissions.

\subsubsection{Scheduling Comparative Analysis with Pre-Filtered Orbital Slots}
While our results from the scheduling comparative analyses show that our concurrent formulation outperforms both baselines on the validation sets, the comparison has some implicit bias, as the placement results of the baselines are subjected to the scheduling constraints from our formulation. To mitigate this bias, another scheduling comparison is performed in which the orbital slots used for the observer placement are pre-filtered. As we have found that communication visibility capabilities are the main restriction on the observer placements selected in the baselines, we use this aspect as the basis for filtering the orbital slots. For the filtering, any orbital slots that can transmit data to the receiver for less than a specified percentage of the mission's time steps are removed. This is intended to filter out orbits that are known to have poor performance due to their inability to alleviate observers' data capacity constraints. The percentage chosen as the threshold for each case is the percentage associated with the orbital slot from the concurrent formulation with the lowest percentage, which ensures that all of the orbital slots initially chosen by the concurrent optimization are still included. This pre-filtering is done for instance~1, and the study is performed again based on the observer placements from the pre-filtered orbital slot set, using the same $20$ validation sets as the prior case study. The statistics of the results for the pre-filtered comparisons are shown in Table~\ref{tab:PP_global_stats}, where the same metrics as the prior comparison are shown.
\begin{table}[htbp]
    \centering
    \caption{Scheduling comparative analysis statistics.}
    \begin{tabular}{l r r r r}
    \hline
    \hline
    Case     &  Mean improvement & Median improvement & Standard deviation & Mean improvement percent, \%\\
    \hline
    Unscheduled & 1680.81 & 2209.21 & 2471.38 & 10.79\\
    Sensor tasking & 2492.71 & 1431.34 & 3567.31 & 121.19\\
    \hline
    \hline
    \end{tabular}
    \label{tab:PP_global_stats}
\end{table}

With the pre-filtering, the concurrent formulation still obtains the better result on average, with mean improvements of $10.79\%$ and $121.19\%$ over the pre-filtered unscheduled and sensor tasking baselines, respectively, and the better result for $15$ and $11$ of the $20$ validation sets. The mean percent improvement over the pre-filtered sensor tasking baseline is dominated by one validation set in which that baseline obtains an objective value of only \num{610.9}; the median percent improvement over this baseline is \num{12.47}\%, and on nine of the $20$ validation sets this baseline obtains the better result, by \num{0.9} to \num{5.9}\%. The pre-filtering thus narrows, but does not close, the gap between the sequential baselines and the concurrent formulation.

\begin{figure}[htbp]
    \centering
    \includegraphics[width=0.9\linewidth]{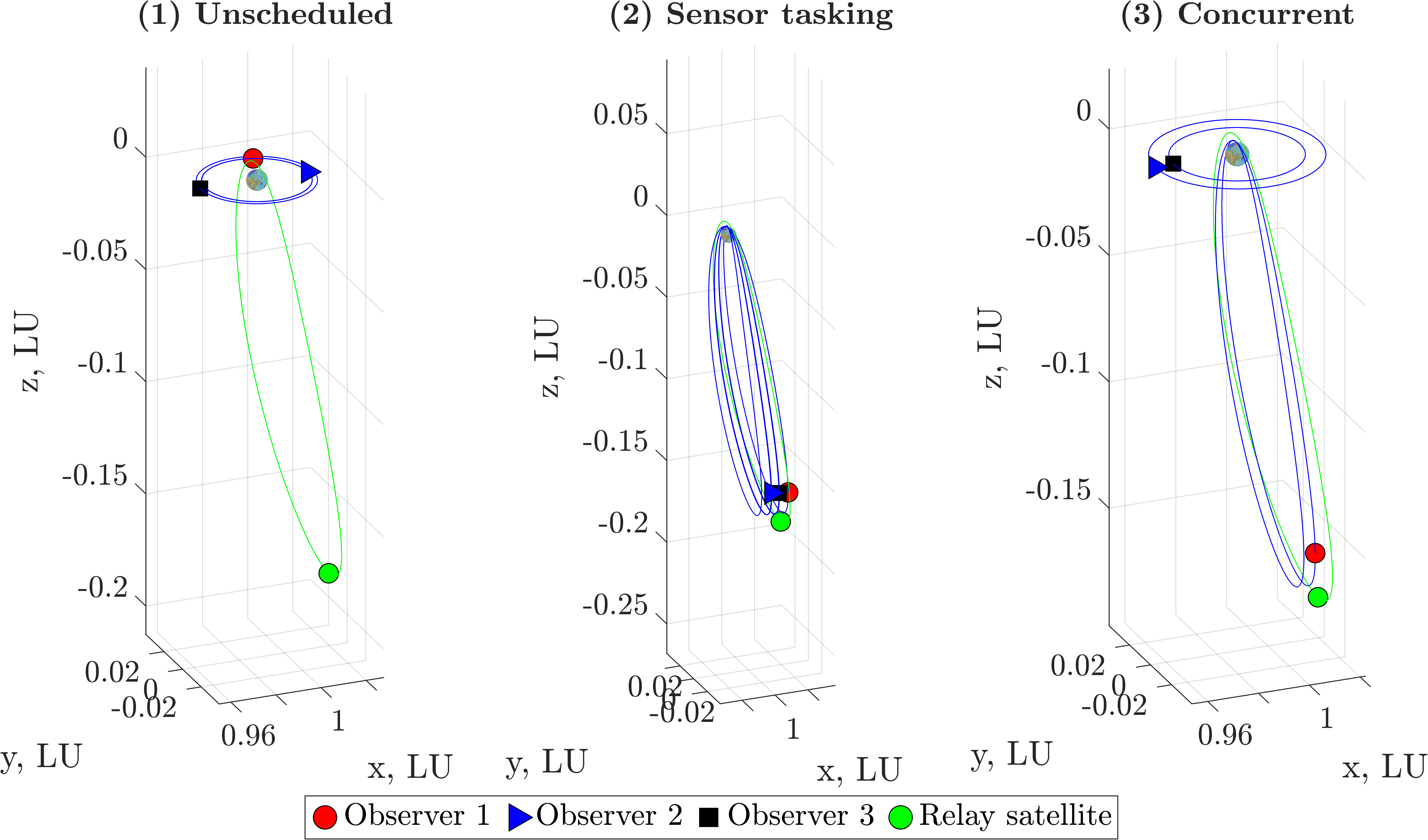}
    \caption{Plots of observer placements for the three pre-filtered cases.}
    \label{fig:PP_placement}
\end{figure}

We again look in more detail at the chosen validation set. A threshold of an orbital slot needing to be able to communicate for at least $29.1\%$ of the time steps is used for this instance, which reduces the number of orbital slots from $135$ to $36$. The placement results of the pre-filtered optimizations are shown in Fig.~\ref{fig:PP_placement}. The placement of the observers in the concurrent formulation has not changed, as the orbital slots chosen by the concurrent formulation in the original optimization remain in the set of orbital slots after the filtering is performed. The pre-filtered unscheduled baseline places all three observers into WLP orbits, one of them in the same orbital slot as the concurrent formulation, while the pre-filtered sensor tasking baseline places all three into SBF orbits, one of them in the same orbital slot as the concurrent formulation. This change in the placement of the observers from the unfiltered cases is a result of the orbital slots chosen prior being filtered out due to their limited ability to communicate with the relay satellite. The initial conditions for the orbital slots chosen by all three cases of the pre-filtered comparative analysis are also provided in Table~\ref{tab:PP_init_cond} in Appendix~B.

Table~\ref{tab:PP_sched_comp_results} reports the same metrics of performance as the unfiltered comparative analysis. The chosen validation set is used again in this comparative analysis. The solutions obtained for this comparative analysis with filtering show improved performance of the baselines in the chosen validation set as compared to their performance without the filtering of the orbital slots.
\begin{table}[htbp]
    \centering
    \caption{Pre-filtered scheduling comparative analysis results.}
    \begin{tabular}{l c c c r r}
    \hline
    \hline
    & \multicolumn{3}{c}{Orbit family}\\
    \cmidrule(r){2-4}
    Case     &  Observer 1 & Observer 2 & Observer 3 & Performance & Improvement over unscheduled, \%\\
    \hline
    Unscheduled & WLP & WLP & WLP & 14455.93 & -\\
    Sensor tasking & SBF & SBF & SBF & 12660.12 & -12.42\\
    Concurrent & SBF & DRO & WLP & 16206.13 & 12.11\\
    \hline
    \hline
    \end{tabular}
    \label{tab:PP_sched_comp_results}
\end{table}

\begin{figure}[htbp]
    \centering
    \begin{subfigure}[b]{0.7\textwidth}
        \centering
        \includegraphics[width=\linewidth]{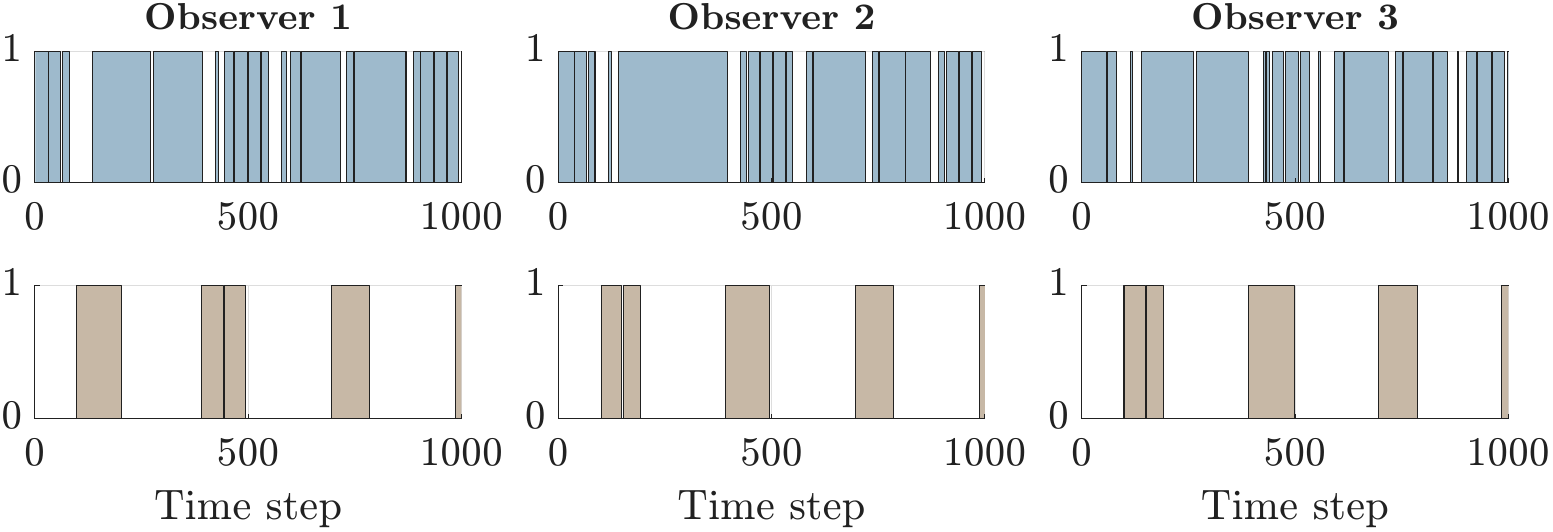}
        \caption{Unscheduled}
	    \label{fig:PP_non_sched_vis}
    \end{subfigure}
    \begin{subfigure}[b]{0.7\textwidth}
        \centering
        \includegraphics[width=\linewidth]{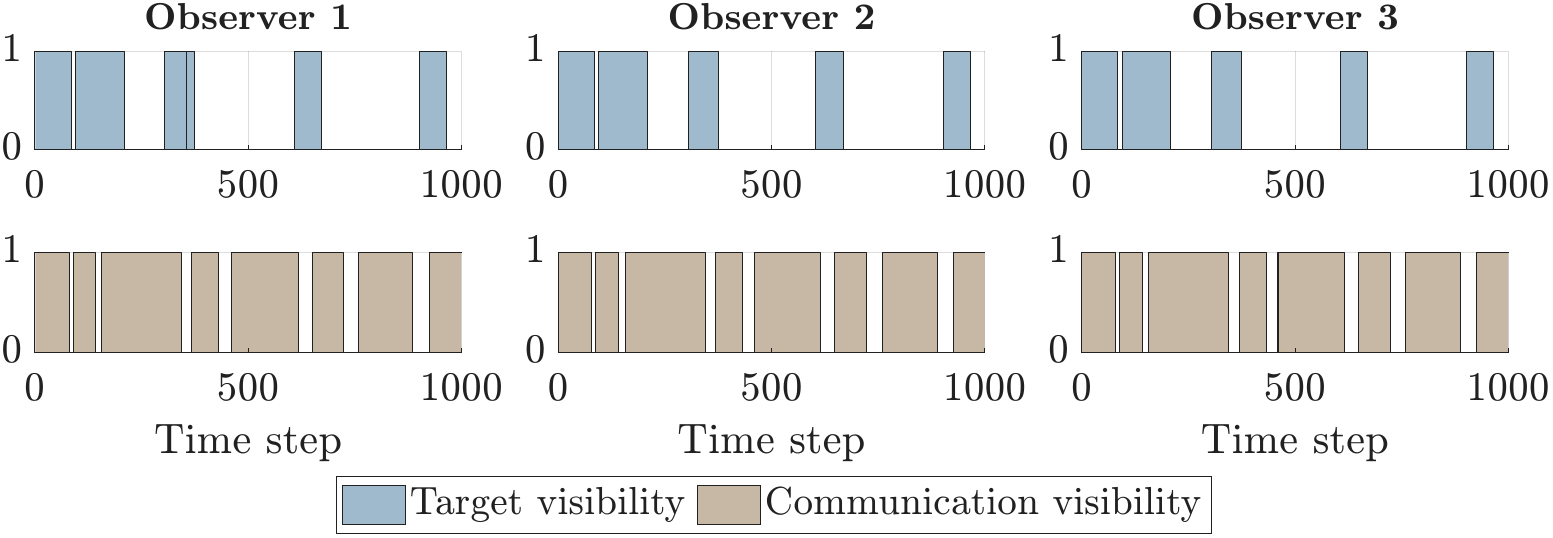}
        \caption{Sensor tasking}
	  \label{fig:PP_sens_task_vis}
    \end{subfigure}
    \caption{Observation and communication visibility of the pre-filtered observers for the chosen validation set.}
\end{figure}

Figures~\ref{fig:PP_non_sched_vis} and \ref{fig:PP_sens_task_vis} show the visibility of the orbital slots chosen by the two pre-filtered baselines. The three WLP observers of the pre-filtered unscheduled baseline see the targets for most of the mission horizon but can communicate with the relay satellite only in three intervals each, while the three SBF observers of the pre-filtered sensor tasking baseline can communicate with the relay satellite for most of the horizon but see the targets only in five intervals. The concurrent formulation combines both: one SBF observer with long communication windows and two observers with long target visibility. In the chosen validation set, this gives it a \num{12.11}\% improvement over the pre-filtered unscheduled baseline and a \num{28.01}\% improvement over the pre-filtered sensor tasking baseline.

\begin{figure}[htbp]
    \centering
    \includegraphics[width=0.7\linewidth]{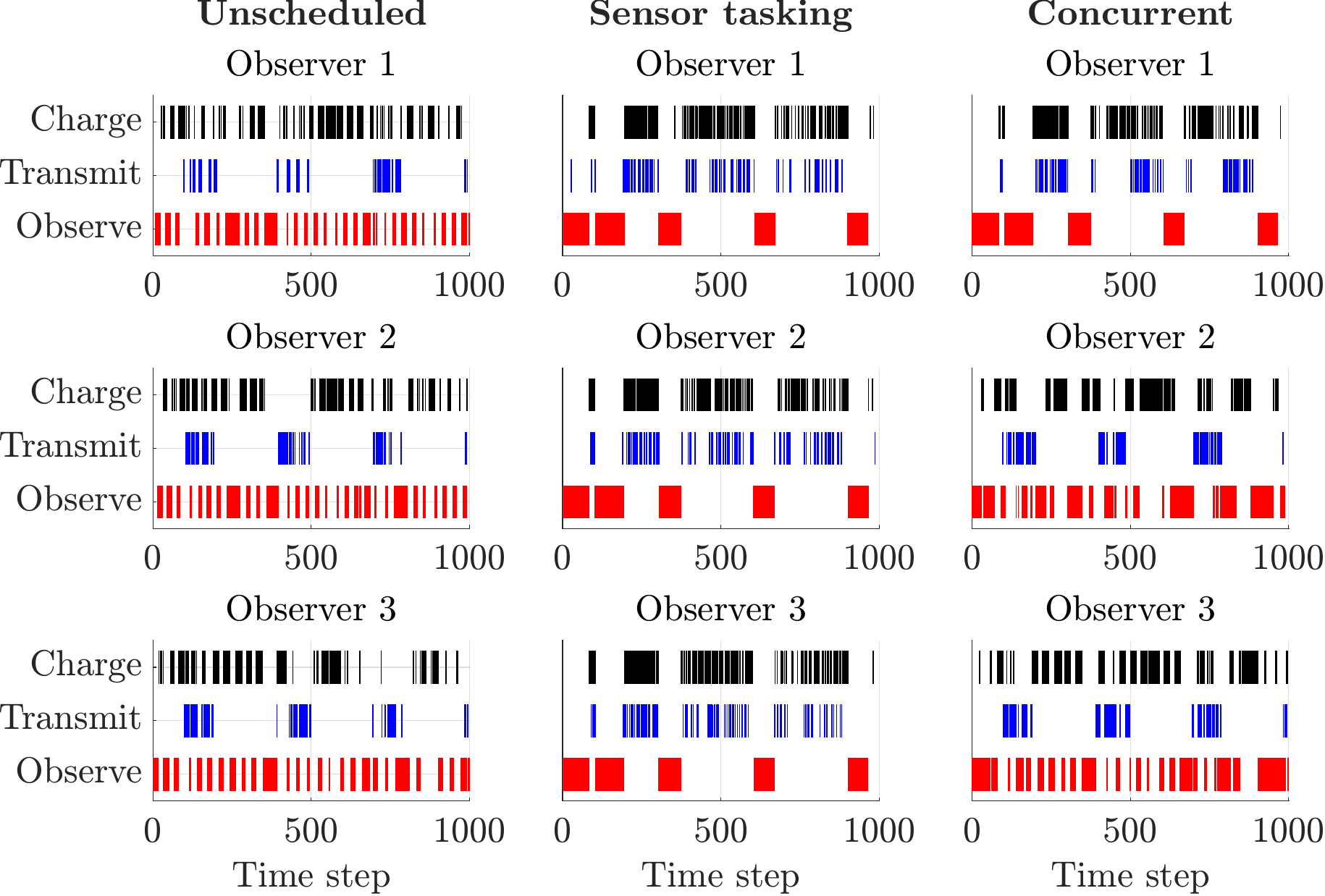}
    \caption{Optimal schedules of the three pre-filtered cases for the chosen validation set.}
    \label{fig:PP_schedules}
\end{figure}

\begin{figure}[htbp]
    \centering
    \includegraphics[width=0.7\linewidth]{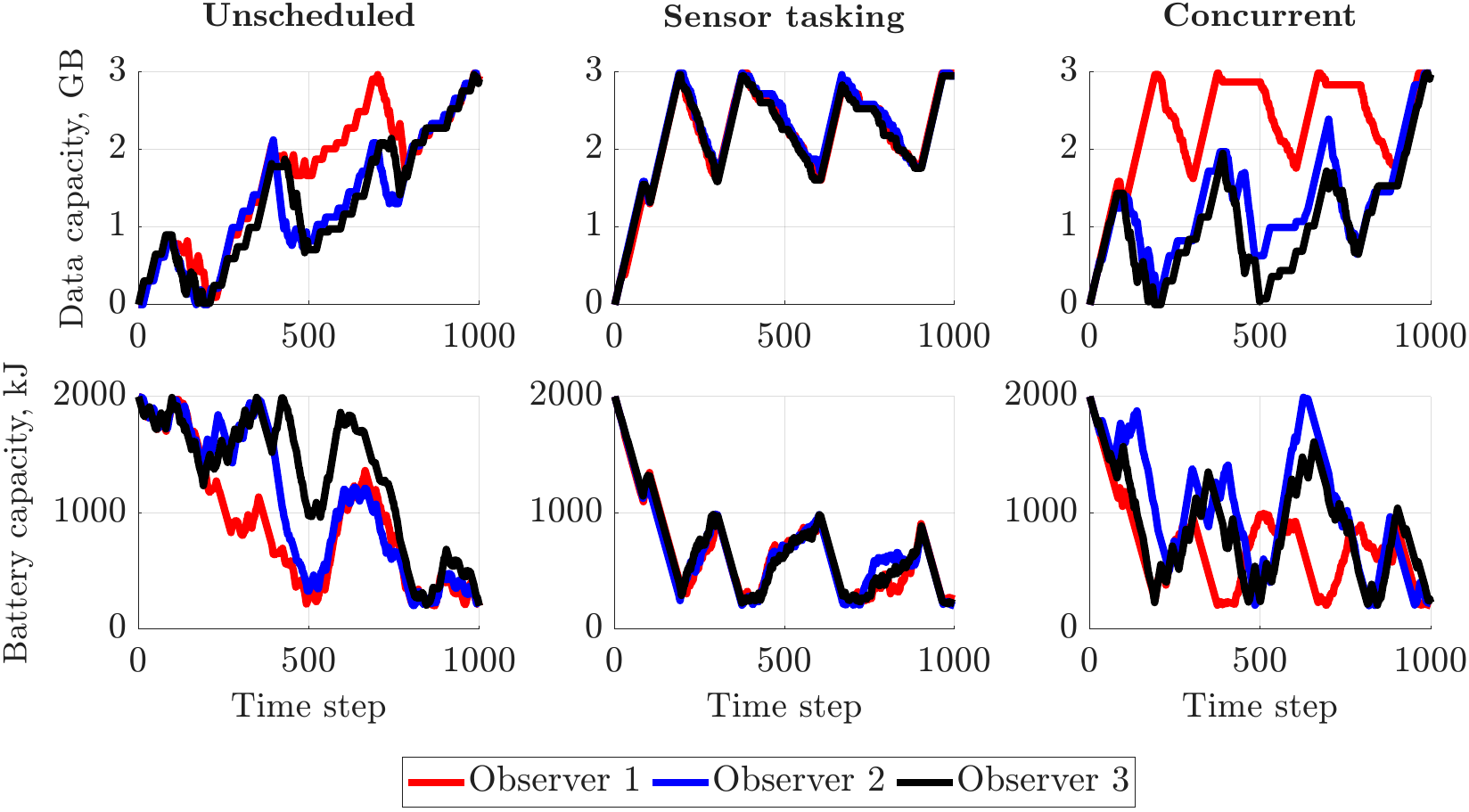}
    \caption{Data and battery capacities of observers during optimal schedules for the three pre-filtered cases.}
    \label{fig:PP_capacities}
\end{figure}

Figure~\ref{fig:PP_schedules} shows the schedules of the observers for the three pre-filtered cases in the chosen validation set, and Fig.~\ref{fig:PP_capacities} shows the corresponding data and battery capacities. With the pre-filtering, the observers of both baselines transmit their data repeatedly, unlike the unfiltered baselines. The observers of the pre-filtered sensor tasking baseline, however, observe only during their few target visibility intervals, and those of the pre-filtered unscheduled baseline accumulate data that they can transmit only in their short communication windows. These results show that filtering the orbital slots by communication capability removes much of the disadvantage of the sequential baselines, but that the placement that balances observation and communication across the observers is found only when the scheduling is incorporated directly into the placement optimization.

\section{Conclusion}
\label{sec:conc}
This paper proposes a concurrent observer placement and scheduling optimization formulation for cislunar SDA. Given multiple scenarios with different sets of targets, the problem aims to maximize the observation performance of a fixed number of space-based observers. The placement of the observers is generalized across all of the scenarios, while the schedules of each observer within the scenarios are different. This scenario-based optimization approach allows us to generalize the placement of the observers for what would be unknown targets, while applying the necessary deterministic scheduling constraints to the problem. The formulation is generalized to allow for any type and amount of cislunar periodic orbits to be considered for the placement. In addition, the data and battery parameters we use can be easily substituted depending on the hardware of the observers being considered.

The MILP formulation for our problem allows users to employ commercial solvers to conveniently and quickly obtain solutions. However, as the scale of the problem grows, commercial solvers struggle to close the optimality gap within a practical runtime. To overcome this issue, we design a column generation approach that is able to find high-quality feasible solutions as the problem grows, including for the largest instance. A comparative analysis is then performed over multiple instances to analyze the performance of the two solution methods for different problem sizes. Gurobi obtains the better solution for every instance with fewer scenarios. For instances with more scenarios, the column generation approach tends to obtain the better solutions wherever the commercial solver struggles to close the MIPGap within the runtime limit. The column generation approach is also able to find a well-performing feasible solution for an instance when Gurobi is not able to find a solution due to memory issues.

To validate the results of our concurrent optimization of placement and scheduling, in which the placement is generalized across scenarios, we conduct a case study for one of the instances from the solution method comparison to test the performance of the placement of the observers in a set of validation sets, each composed of a new set of targets. We then compare these results with the performance of the observer placements from two alternative formulations that perform placement and scheduling optimizations sequentially. We first compare the three cases for the full set of orbital slots, then perform a second comparison in which the orbital slots are filtered based on their communication capabilities. The concurrent formulation shows improvement on average across the set of validation sets for both the unfiltered and filtered comparison cases. These results justify the need for considering the placement and scheduling of observers concurrently for cislunar SDA.

Future research could incorporate the placement and scheduling of the relay satellite into the formulation to create a complete architecture design. Additionally, the use of alternative hardware for the observers could be explored to examine its effects on the placement and scheduling of the observers. An exploration into the effects of the pre-filtering by performing a sweep of multiple thresholds could also be performed. Furthermore, future research could enhance the fidelity of the models used in this work. This could be done by using alternative measurement models, incorporating higher-fidelity constraints on the observers' observation and communication capabilities, or implementing a higher-precision dynamics model for the cislunar domain, such as the Bicircular Restricted Four-Body Problem.

\section*{Appendix A: Full Formulation of Concurrent Placement and Scheduling Optimization}
This Appendix presents the full concurrent placement and scheduling optimization formulation.
\begin{subequations}
\begin{align*}
    \max \quad & \sum_{s \in \mathcal{S}} \sum_{i \in \mathcal{I}} \sum_{t \in \mathcal{T}} \sum_{j \in \mathcal{J}} f_{ijt}^s u_{ijt}^s\\
    \text{s.t.} \quad & \sum_{i \in \mathcal{I}} n_i = N\\
    & V_{ijt}^s n_i \ge u_{ijt}^s, &\quad \forall i \in \mathcal{I}, \forall j \in \mathcal{J}, \forall t \in \mathcal{T}, \forall s \in \mathcal{S}\\
    & Q_{it} n_{i} \ge q_{it}^s, &\quad \forall i \in \mathcal{I}, \forall t \in \mathcal{T}, \forall s \in \mathcal{S}\\
    & W_{it} n_i \ge w_{it}^s, &\quad \forall i \in \mathcal{I}, \forall t \in \mathcal{T}, \forall s \in \mathcal{S}\\
    & \sum_{j \in \mathcal{J}} u_{ijt}^s + q_{it}^s + w_{it}^s \leq 1, &\quad \forall i \in \mathcal{I}, \forall t \in \mathcal{T}, \forall s \in \mathcal{S}\\
    & \sum_{i \in \mathcal{I}} q_{it}^s \leq 1, &\quad \forall t \in \mathcal{T}, \forall s \in \mathcal{S}\\
    & d_{i, t+1}^s = d_{it}^s + D_\text{obs} \sum_{j \in \mathcal{J}}u_{ijt}^s - D_\text{comm}q_{it}^s, &\quad \forall i \in \mathcal{I}, \forall t \in \mathcal{T}\setminus\{T\}, \forall s \in \mathcal{S}\\
    & d_{i, 1}^s = 0, &\quad \forall i \in \mathcal{I}, \forall s \in \mathcal{S}\\
    & b_{i, t+1}^s = b_{it}^s + B_\text{charge} w_{it}^s - B_\text{obs} \sum_{j \in \mathcal{J}} u_{ijt}^s - B_\text{comm} q_{it}^s - B_\text{time} n_i, & \quad \forall i \in \mathcal{I}, \forall t \in \mathcal{T}\setminus\{T\}, \forall s \in \mathcal{S}\\
    & b_{i, 1}^s = B_\text{max}, &\quad \forall i \in \mathcal{I}, \forall s \in \mathcal{S}\\
    & n_i \in \{0,1\}, &\quad \forall i \in \mathcal{I}\\
    & u_{ijt}^s \in \{0,1\}, &\quad \forall i \in \mathcal{I}, \forall j \in \mathcal{J}, \forall t \in \mathcal{T}, \forall s \in \mathcal{S} \\
    & q_{it}^s \in \{0,1\}, &\quad \forall i \in \mathcal{I}, \forall t \in \mathcal{T}, \forall s \in \mathcal{S} \\
    & w_{it}^s \in \{0,1\}, &\quad \forall i \in \mathcal{I}, \forall t \in \mathcal{T}, \forall s \in \mathcal{S} \\
    &d_{it}^s \in [0,D_\text{max}], &\quad \forall i \in \mathcal{I}, \forall t \in \mathcal{T}, \forall s \in \mathcal{S} \\
    &b_{it}^s \in [B_\text{min},B_\text{max}], &\quad \forall i \in \mathcal{I}, \forall t \in \mathcal{T}, \forall s \in \mathcal{S} 
\end{align*}
\end{subequations}

\section*{Appendix B: Initial Conditions of Results from the Scheduling Comparative Analyses}
This Appendix lists the initial conditions of the orbital slots chosen for each of the three cases in both the original and pre-filtered scheduling comparative analyses.
\begin{table}[htb]
    \centering
    \caption{Initial conditions of results from scheduling comparative analysis.}
    \begin{tabular}{l r r r r r r r}
    \hline
    \hline
    Case & Orbit family & $x$, LU & $y$, LU & $z$, LU & $\dot{x}$, LU/TU & $\dot{y}$, LU/TU & $\dot{z}$, LU/TU \\
    \hline
      & L1NH & 0.96631 & 0.09404 & 0.02193 & 0.15271 & -0.18449 & -0.37723\\
     Unscheduled & WLP & 0.90839 & 0.00009 & 0.00000 & -0.00032 & -0.29126 & 0.00000\\
      & WLP & 0.88238 & 0.02274 & 0.00000 & -0.05184 & -0.14744 & 0.00000\\
     \hline
      & NBF & 1.00399 & -0.02018 & 0.15254 & -0.02828 & -0.06327 & -0.15809\\
     Sensor tasking & NBF & 1.00572 & -0.02038 & 0.15268 & -0.02850 & -0.06382 & -0.15761\\
      & L1NH & 0.96631 & 0.09404 & 0.02193 & 0.15271 & -0.18449 & -0.37723\\
     \hline
      & SBF & 1.01004 & -0.02071 & -0.15327 & -0.02868 & -0.06523 & 0.15516\\
     Concurrent & DRO & 0.95565 & 0.00000 & 0.00000 & 0.00000 & 0.64842 & 0.00000\\
      & WLP & 0.96299 & 0.00067 & 0.00000 & -0.01814 & -0.67453 & 0.00000\\
    \hline
    \hline
    \end{tabular}
    \label{tab:init_cond}
\end{table}

\begin{table}[htb]
    \centering
    \caption{Initial conditions of results from pre-filtered scheduling comparative analysis.}
    \begin{tabular}{l r r r r r r r}
    \hline
    \hline
    Case & Orbit family & $x$, LU & $y$, LU & $z$, LU & $\dot{x}$, LU/TU & $\dot{y}$, LU/TU & $\dot{z}$, LU/TU \\
    \hline
      & WLP & 0.99502 & 0.02157 & 0.00000 & -0.67108 & 0.22446 & 0.00000\\
     Unscheduled & WLP & 1.01047 & 0.00000 & 0.00000 & 0.00000 & 0.71138 & 0.00000\\
      & WLP & 0.96299 & 0.00067 & 0.00000 & -0.01814 & -0.67453 & 0.00000\\
     \hline
      & SBF & 1.01548 & -0.02111 & -0.15368 & -0.02872 & -0.06643 & 0.15298\\
     Sensor tasking & SBF & 1.00548 & -0.02032 & -0.15273 & -0.02843 & -0.06380 & 0.15738\\
      & SBF & 1.01004 & -0.02071 & -0.15327 & -0.02868 & -0.06523 & 0.15516\\
     \hline
      & SBF & 1.01004 & -0.02071 & -0.15327 & -0.02868 & -0.06523 & 0.15516\\
     Concurrent & DRO & 0.95565 & 0.00000 & 0.00000 & 0.00000 & 0.64842 & 0.00000\\
      & WLP & 0.96299 & 0.00067 & 0.00000 & -0.01814 & -0.67453 & 0.00000\\
    \hline
    \hline
    \end{tabular}
    \label{tab:PP_init_cond}
\end{table}

\section*{Acknowledgments}
The authors would like to acknowledge Matthew Fox, Brycen Pearl, David Williams Rogers, and Gavin Baker for their continued support throughout this research. Artificial intelligence technologies were used for grammar checking and language refinement. We also thank the anonymous reviewers for their constructive comments, which greatly helped improve the quality of this paper.

\clearpage
\newpage
\bibliography{references}

\end{document}